\documentclass[runningheads]{llncs}

\usepackage[utf8]{inputenc}
\usepackage{amsmath,amssymb,mathtools}
\usepackage[subpreambles=true,mode=buildnew]{standalone}
\usepackage{tikz} 
\usepackage{subcaption}
\usepackage{pdfpages}
\usepackage{hyperref}
\usepackage{import}
\usepackage{wrapfig}
\usepackage{fontawesome}

\newcommand{\myref}[1]{(\ref{#1})}

\usepackage{xargs}
\usepackage{etoolbox} 

\newcommand{\graphname}[3][]{\ensuremath{#2_{#3}^{#1}}}
\newcommand{\chain}[3][]{(\overline{#2}_{#1},{#3},\tau^{\mathcal{#2}}_{#1})}
\newcommandx{\cat}[2][2]{\mbox{\boldmath\ensuremath{\mathsf{#1}}\unboldmath}#2}
\newcommand{\chainname}[2][]{\ensuremath{\mathcal{#2}_{#1}}}
\newcommand{\source}[1][]{\mathit{sc}^{#1}}
\newcommand{\target}[1][]{\mathit{tg}^{#1}}
\newcommand{\elementname}[1]{\texttt{#1}}

\newcommand{\typemorph}[3][]{\ifstrempty{#2}{\ensuremath{\tau^{\mathcal{#1}}}}{\ifstrempty{#3}{\tau_{\mathcal{#2}}^{#1}}{\ensuremath{\tau_{#2,#3}^{\mathcal{#1}}}}}}
\newcommand{\domain}[1]{D(#1)}
\newcommand{\tc}[1]{\overline{#1}}
\newcommand{\partialmap}{{\mbox{\(\hspace{0.7em}\circ\hspace{-1.3em}\longrightarrow\hspace{0.1em}\)}}}

\newcommand{\chainmorph}[2][]{\ensuremath{\sigma_{#1}^{\mathcal{#2}}}}

\newcommand{\indexone}{\ensuremath{i}}
\newcommand{\indextwo}{\ensuremath{j}}
\newcommand{\indexthree}{\ensuremath{k}}
\newcommand{\exampleindexone}{\ensuremath{a}}

\newcommand{\chaindepthone}{\ensuremath{n}}
\newcommand{\chaindepthtwo}{\ensuremath{m}}
\newcommand{\chaindepththree}{\ensuremath{l}}

\newcommand{\examplegraphone}{\ensuremath{G}}
\newcommand{\examplegraphtwo}{\ensuremath{H}}
\newcommand{\examplegraphthree}{\ensuremath{K}}
\newcommand{\examplegraphfour}{\ensuremath{P}}

\newcommand{\graphnodes}{\ensuremath{N}}
\newcommand{\grapharrows}{\ensuremath{A}}
\newcommand{\morphone}{\ensuremath{\phi}}
\newcommand{\morphtwo}{\ensuremath{\psi}}

\newcommand{\maplevelone}{\ensuremath{f}}
\newcommand{\mapleveltwo}{\ensuremath{g}}

\newcommand{\rulegraphleft}{\ensuremath{L}}
\newcommand{\rulegraphmiddle}{\ensuremath{I}}
\newcommand{\rulegraphright}{\ensuremath{R}}
\newcommand{\hierarchygraphleft}{\ensuremath{S}}
\newcommand{\hierarchygraphmiddle}{\ensuremath{D}}
\newcommand{\hierarchygraphright}{\ensuremath{T}}
\newcommand{\rulegraph}{\ensuremath{MM}}
\newcommand{\hierarchygraph}{\ensuremath{TG}}

\newcommand{\maprulelefttomiddle}{\ensuremath{\lambda}}
\newcommand{\maprulerighttomiddle}{\ensuremath{\rho}}
\newcommand{\maphierarchylefttomiddle}{\ensuremath{\varsigma}}
\newcommand{\maphierarchyrighttomiddle}{\ensuremath{\theta}}
\newcommand{\mapruletohierarchybinding}{\ensuremath{\beta}}
\newcommand{\mapruletohierarchyleft}{\ensuremath{\mu}}
\newcommand{\mapruletohierarchymiddle}{\ensuremath{\delta}}
\newcommand{\mapruletohierarchyright}{\ensuremath{\nu}}
\newcommand{\po}{\ensuremath{PO}}
\newcommand{\pb}{\ensuremath{PB}}

\usetikzlibrary{arrows,positioning,decorations.markings,backgrounds}

\definecolor{cdblue}{RGB}{0,0,255}
\definecolor{cdgreen}{RGB}{0,163,0}
\definecolor{cdred}{RGB}{255,0,0}

\tikzset{math/.style={execute at begin node=$, execute at end node=$}}
\tikzset{element/.style={inner sep=2pt,minimum height=1.3em}}
\tikzset{el-math/.style={element,math}}
\tikzset{label/.style={auto,midway,inner sep=2pt}}
\tikzset{la-math/.style={label,math}}
\tikzset{map/.style={->,>=stealth',semithick}}
\tikzset{map2/.style={map,color=black!20}}
\tikzset{mapdots/.style={map,densely dotted}}
\tikzset{incmapl/.style={map,right hook->}}
\tikzset{incmapr/.style={map,left hook->}}
\tikzset{partmap/.style={map,decoration={markings,mark=at position 0.5 with {\draw circle [radius=.4ex];}},postaction={decorate}}}
\tikzset{partmap2/.style={partmap,color=black!20}}
\tikzset{partmapdots/.style={mapdots,decoration={markings,mark=at position 0.5 with {\node[draw,solid,name=s,shape=circle,inner sep=0pt,minimum size=.8ex] {};}},postaction={decorate}}}
\tikzset{mapto/.style={map,|->}}
\tikzset{doublemap/.style={->,double,>=stealth'}}

\begin{document}
\title{Multilevel Typed Graph Transformations}
%

\author{Uwe Wolter\inst{1}\orcidID{0000-0002-7553-9858}\,\textsuperscript{\faEnvelopeO} \and
Fernando Macías\inst{2}\orcidID{0000-0002-6442-6997} \and
Adrian Rutle\inst{3}\orcidID{0000-0002-4158-1644}}
\authorrunning{U.~Wolter \and F.~Macías \and A.~Rutle}
\institute{University of Bergen, Bergen, Norway
\email{Uwe.Wolter@uib.no} 
\and
IMDEA Software Institute, Madrid, Spain
\email{fernando.macias@imdea.org}
\and
Western Norway University of Applied Sciences, Bergen, Norway
\email{aru@hvl.no}}
\maketitle

\begin{abstract}
Multilevel modeling extends traditional modeling techniques with a potentially unlimited number of abstraction levels. 
Multilevel models can be formally represented by multilevel typed graphs whose manipulation and transformation are carried out by multilevel typed graph transformation rules.
These rules are cospans of three graphs and two inclusion graph homomorphisms where the three graphs are multilevel typed over a common typing chain. 
In this paper, we show that typed graph transformations can be appropriately generalized to multilevel typed graph transformations improving preciseness, flexibility and reusability of transformation rules. 
We identify type compatibility conditions, for rules and their matches, formulated as equations and inequations, respectively, between composed partial typing morphisms.
These conditions are crucial presuppositions for the application of a rule for a match---based on a pushout and a final pullback complement construction for the underlying graphs in the category \cat{Graph}---to always provide a well-defined canonical result in the multilevel typed setting. 
Moreover, to formalize and analyze multilevel typing as well as to prove the necessary results, in a systematic way, we introduce the category \cat{Chain} of typing chains and typing chain morphisms.
 
\end{abstract}

\section{Introduction}

Multilevel modeling (MLM) extends conventional techniques from the area of Model-Driven Engineering by providing model hierarchies with multiple levels of abstraction.
The advantages of allowing multiple abstraction levels (e.g. reducing accidental complexity in software models and avoiding synthetic type-instance anti-patterns) and flexible typing (e.g. multiple typing, linguistic extension and deep instantiation), as well as the exact nature of the techniques used for MLM are well studied in the literature~\cite{AtkinsonK01,atkinson2008reducing,atkinson2002rearchitecting,delara2010deep,delara2014whenandhow,dagstuhl2018multi,macias2019thesis}.
Our particular approach \cite{macias2016emisa,macias19jlamp} to MLM 
facilitates the separation of concerns by allowing integration of different multilevel modeling hierarchies as separate aspects of the system to be modelled. 
In addition, we enhance reusability of concepts and their behaviour by allowing the definition of flexible transformation rules which are applicable to different hierarchies with a variable number of levels.
In this paper, we present a revised and extended formalisation of these rules using graph theory and category theory.

\begin{wrapfigure}[15]{l}{.4\textwidth}
  \vspace{-6mm}
  \centering
  \includestandalone[scale=.77]{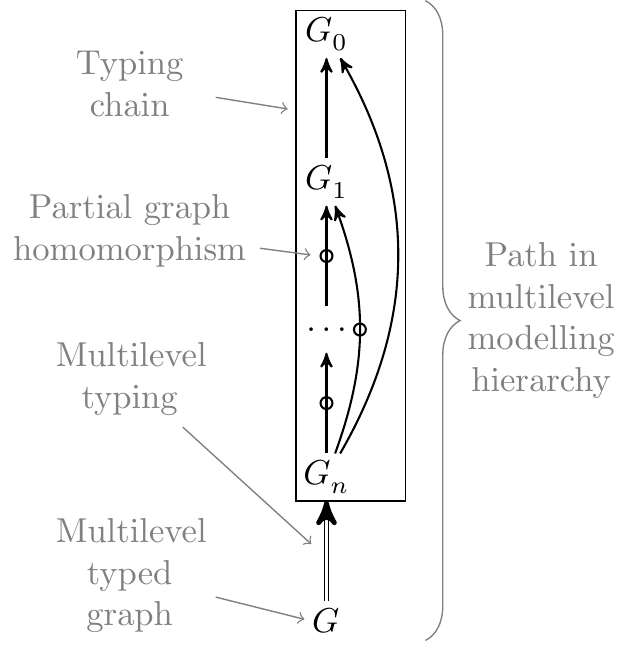}
	\caption{MLM terminology}
	\label{fig:mlm-scheme}
\end{wrapfigure}

As models are usually represented abstractly as graphs, we outline in this paper the graph theoretic foundations of our approach to MLM using multilevel typed graphs, prior to introducing our formalisation of multilevel typed rule definition and application.
Multilevel models are organized in hierarchies, where any graph \(\graphname{\examplegraphone}{}\) is \textit{multilevel typed} over a \textit{typing chain} of graphs (see Fig.~\ref{fig:mlm-scheme}).
The typing relations of elements within each graph are represented via graph morphisms.
Since we allow for deep instantiation~\cite{AtkinsonK01,atkinson2008reducing,atkinson2002rearchitecting,delara2010deep}, which refers to the ability to instantiate an element at any level below the level in which it is defined, these morphisms need to be \textit{partial graph homomorphisms}.
Moreover, more than one model can be typed by the same typing chain (or, conversely, models can be instantiated more than once), hence, all the \textit{paths} that contain such typing relations constitute a full, tree-shaped \textit{multilevel modelling hierarchy} (see Example~\ref{ex:typing-morphisms}).
Finally, the topmost model \(\graphname{\examplegraphone}{0}\) in any hierarchy is fixed, and the typing relations of all models (and the elements inside them) must converge, directly or via a sequence of typing morphisms, into \(\graphname{\examplegraphone}{0}\).
Therefore, the graph morphisms into \(\graphname{\examplegraphone}{0}\) are always total.

Multilevel typed graph transformation rules are cospans 
\begin{tikzpicture}[baseline=-1mm,node distance=13mm]
\node[el-math](l){\graphname{\rulegraphleft}{}};
\node[el-math](i)[right of=l]{\graphname{\rulegraphmiddle}{}};
\node[el-math](r)[right of=i]{\graphname{\rulegraphright}{}};
\draw[incmapl](l)to node[la-math,above](lm){\maprulelefttomiddle}(i);
\draw[incmapr](r)to	node[la-math,above](rm){\maprulerighttomiddle}(i);
\end{tikzpicture} 
of inclusion graph homomorphisms, with \(\rulegraphmiddle=\rulegraphleft\cup\rulegraphright\), where the three graphs are multilevel typed over a common typing chain \chainname{\rulegraph}. 
A match of the left-hand side \graphname{\rulegraphleft}{} of the rule in a graph \graphname{\hierarchygraphleft}{}, at the bottom of a certain hierarchy, multilevel typed over a typing chain \chainname{\hierarchygraph}, is given by a graph homomorphism \(\mapruletohierarchyleft:\graphname{\rulegraphleft}{}\to \graphname{\hierarchygraphleft}{}\) and a flexible typing chain morphism from \chainname{\rulegraph} into \chainname{\hierarchygraph}. The typing chain \chainname{\rulegraph} is local for the rules and is usually different from \chainname{\hierarchygraph} which is determined by the path from \graphname{\hierarchygraphleft}{} to the top of the hierarchy (see Fig.~\ref{fig:mlm-scheme}).

\begin{wrapfigure}[8]{l}{.43\textwidth}
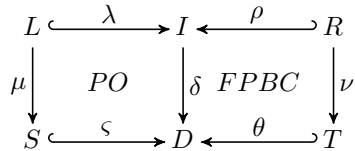

    \vspace{-9mm}
  \centering
  \includestandalone{images/po-pbc-graph-morphisms}
  \vspace{-1ex}
	\caption{Rule structure and basic constructions for rule application}
	\label{fig:basic-constructions-rule-application}
\end{wrapfigure}

To apply these rules we rely on an adaptation of the Sesqui pushout (Sq-PO) approach \cite{CorradiniHHK06} to cospans. 
We construct first the pushout and then the final pullback complement (FPBC) of the underlying graph homomorphisms in the category  \cat{Graph} as shown in Fig.~\ref{fig:basic-constructions-rule-application}. 
Based on these traditional constructions we want to build, in a canonical way, type compatible multilevel typings of the result graphs \graphname{\hierarchygraphmiddle}{} and \graphname{\hierarchygraphright}{} over the  typing chain \chainname{\hierarchygraph}. 
For this to work, we need quite reasonable type compatibility conditions for rules and relatively flexible conditions for matches, formulated as equations and inequations, resp.,  between composed partial typing morphisms.

We introduce typing chain morphisms, and the corresponding category \cat{Chain} of typing chains and typing chain morphisms, to formalize flexible matching and application of multilevel typed rules.
The composition of partial graph homomorphisms is based on pullbacks in the category \cat{Graph}, thus type compatibility conditions can be equivalently expressed by commutativity and pullback conditions in \cat{Graph}.
Therefore, we formalize and analyze multilevel typing as well as describe constructions and prove the intended results, in a systematic way, within the category \cat{Chain}. Especially, we show that the first step in a rule application can be described by a pushout in \cat{Chain}. 
Moreover, the second step is described as a canonical construction in \cat{Chain}, 
however, it is an open question whether this is a final pullback construction in \cat{Chain} or not.

A preliminary version of typing chains are an implicit constituent of the concept ``deep metamodeling stack'' introduced in \cite{rossini2014formalisation} to formalize concepts like parallel linguistic and ontological typing, linguistic extensions, deep instantiation and potencies in deep metamodeling.
We revised this earlier version and further developed it to a concept of its own which serves as a foundation of our approach to multilevel typed model transformations in~\cite{macias19jlamp,wolter2019chains}.
Compared to~\cite{macias19jlamp}, we present in this paper a radically revised and
extended theory of multilevel typed graph transformations. 
In particular, the theory is now more powerful, since we drop the condition 
that typing chain morphisms have to be closed (see Def.~\ref{def:graph-chain-morphism}). 
Moreover, we detail the FPBC step which is missing in \cite{macias19jlamp}.
Due to space limitations, we will not present the background results concerning the equivalence between the practice of individual direct typing -- which are used in applications and implementations -- and our categorical reformulation of this practice by means of typing chains. 
These equivalence 
results as well as examples and proofs can be found
in~\cite{wolter2019chains}.

\section{Typing Chains and Multilevel Typing of Graphs}
\label{sec:typing-chains}

\cat{Graph} denotes the category of (directed multi-) graphs \(\graphname{\examplegraphone}{} = {(\graphname[\graphnodes]{\examplegraphone}{}, \graphname[\grapharrows]{\examplegraphone}{}, \source[\examplegraphone], \target[\examplegraphone])}\) and graph homomorphisms \(\morphone=(\morphone^\graphnodes,\morphone^\grapharrows):\graphname{\examplegraphone}{}\to\graphname{\examplegraphtwo}{}\) \cite{ehrig2006fundamentals}. We will use the term \textbf{element} to refer to both nodes and arrows.

Multilevel typed graphs are graphs typed over a typing chain, i.e., a sequence \([\graphname{\examplegraphone}{\chaindepthone}, \graphname{\examplegraphone}{\chaindepthone-1}, \dots, \graphname{\examplegraphone}{1}, \graphname{\examplegraphone}{0}]\) of graphs where the elements in any of the graphs \(\graphname{\examplegraphone}{\indexone}\), with \(\chaindepthone\geq\indexone \geq 1\),  are, on their part,  multilevel typed over the sequence \([\graphname{\examplegraphone}{\indexone-1}, \dots, \graphname{\examplegraphone}{1}, \graphname{\examplegraphone}{0}]\). Paths in our MLM hierarchies give rise to typing chains. 
The indexes \(\indexone\) refer to the abstraction levels in a modeling hierarchy where \(0\)  denotes the most abstract top level.

Following well-established approaches in the Graph Transformations field \cite{ehrig2006fundamentals}, we define typing  by means of graph homomorphisms. 
This enables us to establish and develop our approach by reusing, variating, and extending the wide range of constructions and results achieved in that field.
Moreover, this paves the way to generalize  the present ``paradigmatic'' approach, where models are just graphs, to more sophisticated kinds of diagrammatic models, especially those that take advantage of diagrammatic constraints \cite{rossini2014formalisation,rutle2012formal}. 

We allow typing to jump over abstraction levels, i.e., an element 
in graph \(\graphname{\examplegraphone}{\indexone}\) may have no type in \(\graphname{\examplegraphone}{\indexone-1}\) but only in one (or more) of the graphs \(\graphname{\examplegraphone}{\indexone-2}, \dots, \graphname{\examplegraphone}{1}, \graphname{\examplegraphone}{0}\).
Two different elements in the same graph may have their types located in different graphs along the typing chain.
To formalize this kind of flexible typing, we use partial graph homomorphisms that we introduced already in~\cite{rossini2014formalisation}. 
\begin{definition}
	A \textbf{partial graph homomorphism} \(\varphi: \graphname{\examplegraphone}{} \partialmap \graphname{\examplegraphtwo}{}\) 
	is given by a subgraph \(\domain{\varphi} \sqsubseteq \graphname{\examplegraphone}{}\), called the \textbf{domain of definition} of \(\varphi\), and a graph homomorphism \(\varphi: \domain{\varphi} \xrightarrow{} \graphname{\examplegraphtwo}{} \).
\end{definition}
Note that we use, in abuse of notation, the same name for both the partial and the corresponding total graph homomorphisms.
%
To express transitivity of typing and later also compatibility of typing, we need as well the composition of partial graph homomorphisms as a partial order between partial graph homomorphisms.
\begin{definition}
	\label{def:composition-partial-graph-homomorphisms}
	The \textbf{composition} \(\varphi;\psi:\graphname{\examplegraphone}{} \partialmap \graphname{\examplegraphthree}{}\) of two partial graph homomorphisms \(\varphi: \graphname{\examplegraphone}{} \partialmap \graphname{\examplegraphtwo}{}\) and \(\psi:\graphname{\examplegraphtwo}{} \partialmap \graphname{\examplegraphthree}{}\) is defined as follows:
	\begin{itemize}
		\item \(\domain{\varphi;\psi} := \varphi^{-1} (\domain{\psi})\), 
		\item \((\varphi;\psi)^\graphnodes(\elementname{e}):= \psi^\graphnodes(\varphi^\graphnodes(\elementname{e}))\) for all 
		\(\elementname{e}\in\domain{\varphi;\psi}^\graphnodes\) and \((\varphi;\psi)^\grapharrows(\elementname{f}):= \psi^\grapharrows(\varphi^\grapharrows(\elementname{f}))\) for all 
		\(\elementname{f}\in\domain{\varphi;\psi}^\grapharrows\).
	\end{itemize}
	More abstractly, the composition of two partial graph homomorphisms is defined by the following commutative diagram of total graph homomorphisms. 
	\begin{center}
		\begin{tikzpicture}[on grid,node distance=20mm]
		
		\def\vd{7mm}
		\def\hd{20mm}
		
		\node[el-math] (dtki)										{\domain{\varphi;\psi}};
		\node[el-math] (dtkj)	[below left=\vd and \hd of dtki]	{\domain{\varphi}};
		\node[el-math] (dtji)	[below right=\vd and \hd of dtki]	{\domain{\psi}};
		\node[el-math] (gk)		[below left=\vd and \hd of dtkj]	{\graphname{\examplegraphone}{}};
		\node[el-math] (gj)		[below right=\vd and \hd of dtkj]	{\graphname{\examplegraphtwo}{}};
		\node[el-math] (gi)		[below right=\vd and \hd of dtji]	{\graphname{\examplegraphthree}{}};
		
		\draw[incmapr]			(dtki) to node[la-math,above left]	(in1)   {\sqsubseteq}		(dtkj);
		\draw[map]				(dtki) to node[la-math,above]   	(map1)	{\varphi_{|\psi}}	(dtji);
		\draw[incmapr]			(dtkj) to node[la-math,above left]	(in2)	{\sqsubseteq}		(gk);
		\draw[map]				(dtkj) to node[la-math,above]		(tkj)	{\varphi}			(gj);
		\draw[incmapr]			(dtji) to node[la-math,above left]	(in3)	{\sqsubseteq}		(gj);
		\draw[map]				(dtji) to node[la-math,above]		(tji)	{\psi}				(gi);
		\draw[map,bend left]	(dtki) to node[la-math,above right]	(tki)	{\varphi;\psi}		(gi);
		
		\node[el-math] (pb)		[below=\vd and of dtki]				{\pb};
		\end{tikzpicture}
	\end{center}
	Note that \(\domain{\varphi;\psi} = \domain{\varphi}\) if \(\psi\) is total, i.e., \(\graphname{\examplegraphtwo}{} = \domain{\psi}\).
\end{definition}

%
\begin{definition}
	\label{def:order-partial-graph-homomorphisms}
	For any two partial graph homomorphisms \(\varphi,\phi: \graphname{\examplegraphone}{} \partialmap \graphname{\examplegraphtwo}{}\) 
	we have 
	\(\varphi \preceq \phi\) iff \(\domain{\varphi} \sqsubseteq \domain{\phi}\) and
	\(\varphi\), \(\phi\) coincide on \(\domain{\varphi}\). 
		
		
		
		
\end{definition}

Now, we can define typing chains as a foundation for our investigation of multilevel typed graph transformations in the rest of the paper.
\begin{definition}
	\label{def:typing-chain}
	A \textbf{typing chain} \(\chainname{\examplegraphone} = \chain{\examplegraphone}{\chaindepthone}\) is given by a natural number \(\chaindepthone\), a sequence \(\tc{\examplegraphone} = [\graphname{\examplegraphone}{\chaindepthone}, \graphname{\examplegraphone}{\chaindepthone-1}, \dots, \graphname{\examplegraphone}{1}, \graphname{\examplegraphone}{0}]\) of graphs of length \(\chaindepthone+1\)
	and a family \(\typemorph[\examplegraphone]{}{} = (\typemorph[\examplegraphone]{\indextwo}{\indexone} : \graphname{\examplegraphone}{\indextwo} \partialmap \graphname{\examplegraphone}{\indexone} \mid \chaindepthone \geq \indextwo >\indexone \geq 0 )\) of partial graph homomorphisms, called \textbf{typing morphisms}, satisfying the following properties:
	\begin{itemize}
		\item \textbf{Total:} All the morphisms   
		\(\typemorph[\examplegraphone]{\indextwo}{0} : \graphname{\examplegraphone}{\indextwo} \to \graphname{\examplegraphone}{0}\) with \(\chaindepthone \geq \indextwo \geq 1\) are total.
		\item \textbf{Transitive:} For all 
		\(\chaindepthone\geq\indexthree > \indextwo > \indexone \geq 0\) we have \;\(\typemorph[\examplegraphone]{\indexthree}{\indextwo};\typemorph[\examplegraphone]{\indextwo}{\indexone} \preceq \typemorph[\examplegraphone]{\indexthree}{\indexone}\).
		\item \textbf{Connex:} For all 
		\(\chaindepthone\geq\indexthree > \indextwo > \indexone \geq 0\) we have \(\domain{\typemorph[\examplegraphone]{\indexthree}{\indextwo}}  \cap\domain{\typemorph[\examplegraphone]{\indexthree}{\indexone}} \sqsubseteq \domain{\typemorph[\examplegraphone]{\indexthree}{\indextwo};\typemorph[\examplegraphone]{\indextwo}{\indexone}}
		=(\typemorph[\examplegraphone]{\indexthree}{\indextwo})^{-1} (\domain{\typemorph[\examplegraphone]{\indextwo}{\indexone}})
		\), moreover, \(\typemorph[\examplegraphone]{\indexthree}{\indextwo};\typemorph[\examplegraphone]{\indextwo}{\indexone}\) and \(\typemorph[\examplegraphone]{\indexthree}{\indexone}\) coincide on \(\domain{\typemorph[\examplegraphone]{\indexthree}{\indextwo}}  \cap\domain{\typemorph[\examplegraphone]{\indexthree}{\indexone}}\).
	\end{itemize}
	Due to Definitions \ref{def:composition-partial-graph-homomorphisms} and \ref{def:order-partial-graph-homomorphisms}, transitivity and connexity together mean that 
	\(\domain{\typemorph[\examplegraphone]{\indexthree}{\indextwo}}  \cap\domain{\typemorph[\examplegraphone]{\indexthree}{\indexone}} = \domain{\typemorph[\examplegraphone]{\indexthree}{\indextwo};\typemorph[\examplegraphone]{\indextwo}{\indexone}}\),
	i.e., we do have a (unique) total graph homomorphism \(\typemorph[\examplegraphone]{\indexthree}{\indextwo|\indexone}:
		\domain{\typemorph[\examplegraphone]{\indexthree}{\indextwo}}\cap\domain{\typemorph[\examplegraphone]{\indexthree}{\indexone}} \to \domain{\typemorph[\examplegraphone]{\indextwo}{\indexone}}\) and the following commutative diagram of total graph homomorphisms
		\vspace{-2ex}
		\begin{center}
			\begin{tikzpicture}[on grid,node distance=40mm]
			
			\def\vd{13mm}
			
			\node[el-math]	(dki)						{\domain{\typemorph[\examplegraphone]{\indexthree}{\indexone}}};
			\node[el-math]	(gi)	[right of=dki]		{\graphname{\examplegraphone}{\indexone}};
			\node[el-math]	(dkji)	[below=\vd of dki]	{\domain{\typemorph[\examplegraphone]{\indexthree}{\indextwo}} \cap\domain{\typemorph[\examplegraphone]{\indexthree}{\indexone}}};
			\node[el-math]	(dji)	[right of=dkji]		{\domain{\typemorph[\examplegraphone]{\indextwo}{\indexone}}};
			\node[el-math]	(dkj)	[below=\vd of dkji]	{\domain{\typemorph[\examplegraphone]{\indexthree}{\indextwo}}};
			\node[el-math]	(gj)	[right of=dkj]		{\graphname{\examplegraphone}{\indextwo}};
			\node[el-math]	(gk)	[left=30mm of dkji]	{\graphname{\examplegraphone}{\indexthree}};
			
			\draw[map]				(dki)	to	node [la-math]				(tki)	{\typemorph[\examplegraphone]{\indexthree}{\indexone}}			(gi);
			\draw[map]				(dkji)	to	node [la-math]				(tkji)	{\typemorph[\examplegraphone]{\indexthree}{\indextwo|\indexone}}(dji);
			\draw[map]				(dkj)	to	node [la-math]				(tkj)	{\typemorph[\examplegraphone]{\indexthree}{\indextwo}}			(gj);
			\draw[incmapr]			(dkji)	to	node [la-math,right]		(indki)	{\sqsubseteq}													(dki);
			\draw[incmapl]			(dkji)	to	node [la-math]				(indkj)	{\sqsubseteq}													(dkj);
			\draw[map]				(dji)	to	node [la-math,right]		(djigi)	{\typemorph[\examplegraphone]{\indextwo}{\indexone}}			(gi);
			\draw[incmapr]			(dji)	to	node [la-math,right]		(ingj)	{\sqsubseteq}													(gj);
			\draw[incmapl,bend right](dki)	to	node [la-math,above left]	(ingk1)	{\sqsubseteq}													(gk);
			\draw[incmapr,bend left](dkj)	to	node [la-math]				(ingk2)	{\sqsubseteq}													(gk);
			
			\node	(comm1)	[below right=6mm and 19mm of dki]	{=};
			\node	(pb1)	[left=19mm of dkji]					{\pb};
			\node	(pb2)	[below right=6mm and 19mm of dkji]	{\pb}; 	
			\end{tikzpicture}
		\end{center}
\end{definition}
\begin{remark}
\label{rem:direct-types}
For any element \elementname{e} in any graph \graphname{\examplegraphone}{\indexone} in a typing chain, with \(\indexone>0\), there exists a unique index \(m_\elementname{e}\), with \(\indexone>m_\elementname{e}\geq 0\), such
that \elementname{e} is in the domain of the typing morphism \typemorph[\examplegraphone]{\indexone}{m_\elementname{e}} but not in the domain of any typing morphism \typemorph[\examplegraphone]{\indexone}{\indextwo} with \( \indexone>\indextwo >m_\elementname{e}\). We call \(\typemorph[\examplegraphone]{\indexone}{m_\elementname{e}}(\elementname{e})\) the \textbf{direct type} of \elementname{e}. For any other index \indexthree, with \(m_\elementname{e}>\indexthree \geq 0\), we call \(\typemorph[\examplegraphone]{\indexone}{\indexthree}(\elementname{e})\), if it is defined, a \textbf{transitive type} of \elementname{e}.
\end{remark}

\begin{figure}
		\centering
		\includestandalone[width=.6\linewidth]{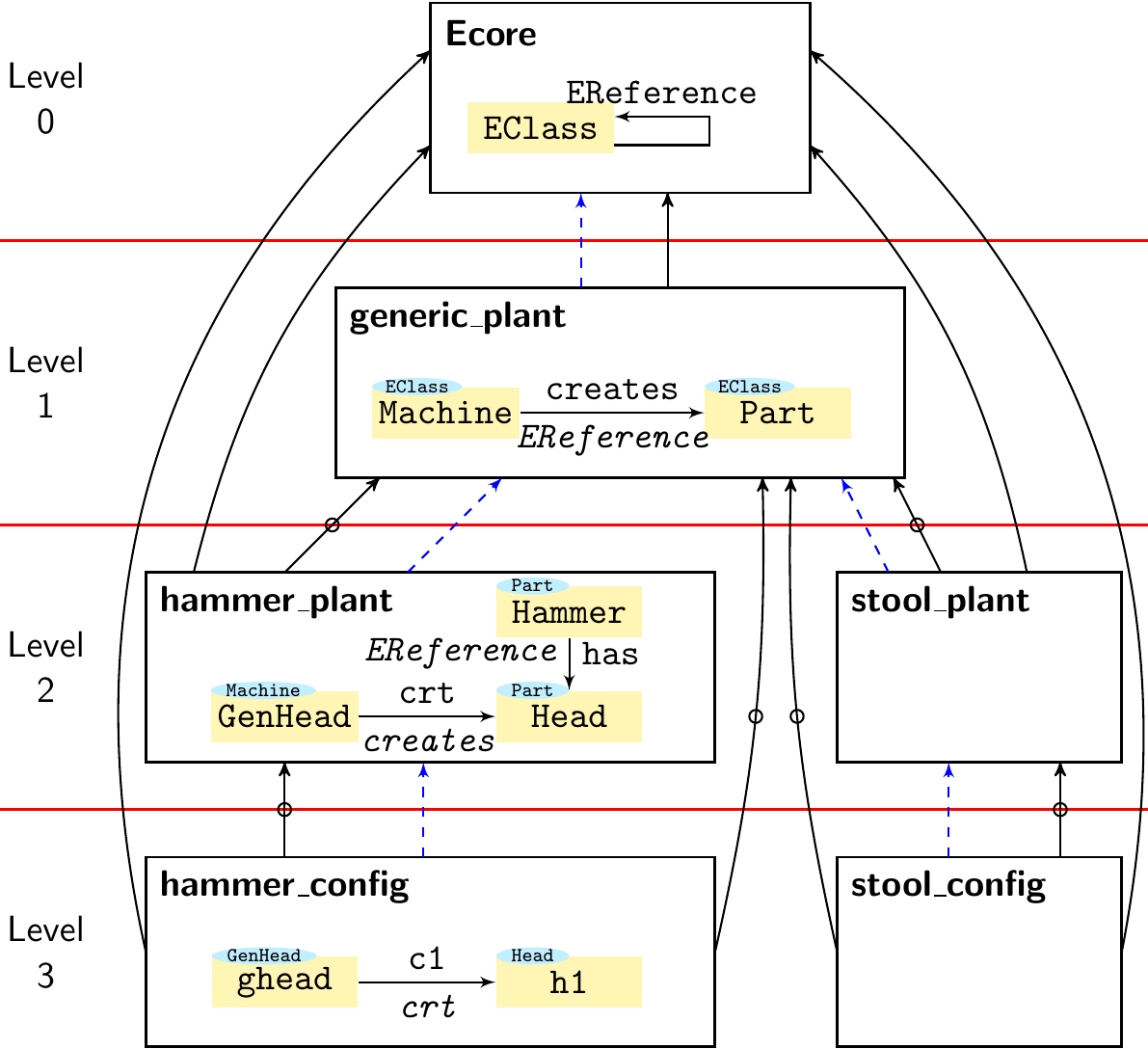}
		\caption{Multilevel modeling hierarchy with typing morphisms}
		\label{fig:hierarchy-introductory-example-4}
	\end{figure}

\begin{example}
	\label{ex:typing-morphisms}
	Fig.~\ref{fig:hierarchy-introductory-example-4} depicts the typing morphisms between the graphs in a simplified sample hierarchy. The direct types for nodes and arrows are indicated with blue and cursive labels, respectively. All  typing morphisms in the simple typing chain \chainname{TG}, determined by the sequence
	\([\elementname{hammer\_plant},\elementname{generic\_plant},\elementname{Ecore}]
	\) of graphs,
	 are total except the one from \elementname{hammer\_plant} to \elementname{generic\_plant}, since the direct type of \elementname{has} is located in \elementname{Ecore}. 
    We have chosen Ecore as the top-most graph since it 
    provides implementation support through the Eclipse Modeling Framework~\cite{steinberg2008emf}. 
    This enables our approach to MLM to exploit the best from fixed-level and multi-level concepts~\cite{macias2016multecore}.
    \qed
\end{example}

To describe later the flexible matching of multilevel typed rules and the result of rule applications, we need a corresponding flexible notion of morphisms between typing chains.

\begin{figure}[ht!]
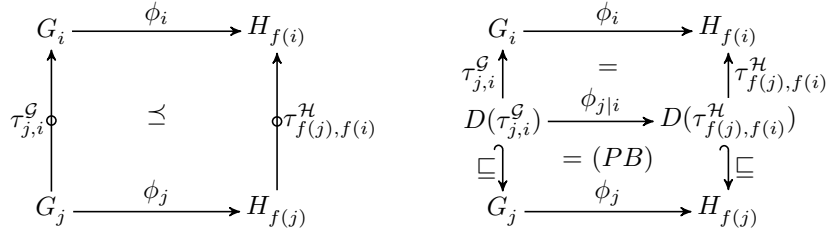

	\begin{center}
		\includestandalone{images/mlm-binding-diagram-formal}
		\vspace{-1ex}
		\caption{Establishing a morphism between two typing chains, level-wise}
		\label{fig:mlm-binding-diagram-formal}
	\end{center}
	\vspace{-4ex}	
\end{figure}

\begin{definition}
	\label{def:graph-chain-morphism}
	A \textbf{typing chain morphism} \((\morphone,\maplevelone) : \chainname{\examplegraphone} \to \chainname{\examplegraphtwo}\) between two typing chains \(\chainname{\examplegraphone} = \chain{\examplegraphone}{\chaindepthone}\) and \(\chainname{\examplegraphtwo} = \chain{\examplegraphtwo}{\chaindepthtwo}\) with \(\chaindepthone \leq \chaindepthtwo\) is given by
	\begin{itemize}
		\item a function \(\maplevelone : [\chaindepthone] \to [\chaindepthtwo] \), where \([\chaindepthone] = \{0, 1, 2, \ldots, \chaindepthone\} \), such that (1) \(\maplevelone(0) = 0\) and (2) \(\indextwo > \indexone\) implies 
		\(\maplevelone(\indextwo) - \maplevelone(\indexone)\geq \indextwo - \indexone\)  for all \(\indexone, \indextwo \in [\chaindepthone]\), and
		\item a family of total graph homomorphisms \(\morphone = (\morphone_\indexone : \graphname{\examplegraphone}{\indexone} \to \graphname{\examplegraphtwo}{\maplevelone(\indexone)} \mid \indexone \in [\chaindepthone] )\) such that
		\begin{equation}\label{eq:typing-chain-morphism}
		\typemorph[\examplegraphone]{\indextwo}{\indexone} ; \morphone_\indexone \preceq \morphone_\indextwo ; \typemorph[\examplegraphtwo]{\maplevelone(\indextwo)}{\maplevelone(\indexone)}\quad \mbox{for all} \; \chaindepthone \geq \indextwo > \indexone \geq 0,
		\end{equation}
		i.e., due to Definitions~\ref{def:composition-partial-graph-homomorphisms} and \ref{def:order-partial-graph-homomorphisms}, we assume for any \(\chaindepthone \geq \indextwo > \indexone \geq 0\) the existence of a total graph homomorphism \(\phi_{\indextwo \mid \indexone}\) that makes the diagram of total graph homomorphisms displayed in Fig.~\ref{fig:mlm-binding-diagram-formal} commutative.
	\end{itemize}
	A typing chain morphism \((\morphone,\maplevelone) : \chainname{\examplegraphone} \to \chainname{\examplegraphtwo}\) is \textbf{closed} iff \(\typemorph[\examplegraphone]{\indextwo}{\indexone} ; \morphone_\indexone = \morphone_\indextwo ; \typemorph[\examplegraphtwo]{\maplevelone(\indextwo)}{\maplevelone(\indexone)}\) for all \(\chaindepthone \geq \indextwo > \indexone \geq 0\), i.e.,  the right lower square in Fig.~\ref{fig:mlm-binding-diagram-formal} is a pullback.
\end{definition}
%


Typing morphisms are composed by the composition of commutative squares.
\begin{definition}
	\label{def:composition-graph-chain-morphisms}
	The \textbf{composition} \((\morphone,\maplevelone);(\morphtwo,\mapleveltwo) : \chainname{\examplegraphone} \to \chainname{\examplegraphthree}\) of two  typing chain morphisms \((\morphone,\maplevelone) : \chainname{\examplegraphone} \to \chainname{\examplegraphtwo}\), \((\morphtwo,\mapleveltwo) : \chainname{\examplegraphtwo} \to \chainname{\examplegraphthree}\) between typing chains \(\chainname{\examplegraphone} = \chain{\examplegraphone}{\chaindepthone}\), \(\chainname{\examplegraphtwo} = \chain{\examplegraphtwo}{\chaindepthtwo}\), \(\chainname{\examplegraphthree} = \chain{\examplegraphthree}{\chaindepththree}\) with \(\chaindepthone \leq \chaindepthtwo \leq \chaindepththree\) is defined by \;
	\((\morphone,\maplevelone);(\morphtwo,\mapleveltwo) := (\morphone;\morphtwo_{\downarrow \maplevelone},\maplevelone;\mapleveltwo)\), 
	where \(\morphtwo_{\downarrow \maplevelone} := (\morphtwo_{\maplevelone(\indexone)} : \graphname{\examplegraphtwo}{\maplevelone(\indexone)} \to \graphname{\examplegraphthree}{\mapleveltwo(\maplevelone(\indexone))} \mid \indexone \in [\chaindepthone])\), and thus
	\(\morphone;\morphtwo_{\downarrow \maplevelone} := (\morphone_\indexone;\morphtwo_{\maplevelone(\indexone)} : \graphname{\examplegraphone}{\indexone} \to \graphname{\examplegraphthree}{\mapleveltwo(\maplevelone(\indexone))} \mid \indexone \in [\chaindepthone])\).
\end{definition}

\cat{Chain} denotes the category of typing chains and typing chain morphisms.
%
%

A natural way to define multilevel typing of a graph \graphname{\examplegraphtwo}{} over a typing chain  \(\chainname{\examplegraphone}\) would be a family \(\chainmorph{} = (\chainmorph[\indexone]{} : \graphname{\examplegraphtwo}{} \partialmap \graphname{\examplegraphone}{\indexone} \mid \chaindepthone\geq\indexone\geq 0)\) of partial graph homomorphisms satisfying certain properties. 
However, as shown in~\cite{wolter2019chains}, those families are not appropriate to state adequate type compatibility requirements for rules and matches and to construct the results of rule applications. 
Therefore, we employ the sequence of the domains of definition of the \chainmorph[\indexone]{}'s as a typing chain and describe multilevel typing by means of typing chain morphisms. 
The following lemma describes how any sequence of subgraphs gives rise to a typing chain.
\begin{lemma}
\label{lem:inclusion-chain}
	Any sequence \(\tc{\examplegraphtwo} = [\graphname{\examplegraphtwo}{\chaindepthone}, \graphname{\examplegraphtwo}{\chaindepthone-1}, \dots, \graphname{\examplegraphtwo}{1}, \graphname{\examplegraphtwo}{0}]\) of subgraphs of a graph \(\graphname{\examplegraphtwo}{}\), with \(\graphname{\examplegraphtwo}{0} = \examplegraphtwo\), can be extended to a typing chain \(\chainname{\examplegraphtwo} = \chain{\examplegraphtwo}{\chaindepthone}\) where for all \(\chaindepthone \geq \indextwo >\indexone \geq 0\) the corresponding \textbf{typing morphism}  \(\typemorph[\examplegraphtwo]{\indextwo}{\indexone} : \graphname{\examplegraphtwo}{\indextwo} \partialmap \graphname{\examplegraphtwo}{\indexone}\) is given by \(\domain{\typemorph[\examplegraphtwo]{\indextwo}{\indexone}} := \graphname{\examplegraphtwo}{\indextwo} \cap \graphname{\examplegraphtwo}{\indexone}\) and the span of total inclusion graph homomorphisms
	\vspace{-1.5ex}
	\begin{center}
		\begin{tikzpicture}[baseline=-1mm,node distance=30mm]
		\node[el-math](hj){\graphname{\examplegraphtwo}{\indextwo}};
		\node[el-math](dji)[right of=hj]{\domain{\typemorph[\examplegraphtwo]{\indextwo}{\indexone}} = \graphname{\examplegraphtwo}{\indextwo} \cap \graphname{\examplegraphtwo}{\indexone}};
		\node[el-math](hi)[right of=dji]{\graphname{\examplegraphtwo}{\indexone}};
		\draw[incmapr](dji)to node[la-math,above](lm){\sqsubseteq}(hj);
		\draw[incmapl](dji)to node[la-math,above](rm){\typemorph[\examplegraphtwo]{\indextwo}{\indexone}}(hi);
		\end{tikzpicture} .
		\vspace{-.5ex}
	\end{center} 

	We call the typing chain \(\chainname{\examplegraphtwo} = \chain{\examplegraphtwo}{\chaindepthone}\)  an \textbf{inclusion chain on \graphname{\examplegraphtwo}{}}.
\end{lemma}
A \textbf{multilevel typing of a graph} \graphname{\examplegraphtwo}{} over a typing chain \(\chainname{\examplegraphone} = \chain{\examplegraphone}{\chaindepthone}\) is given by an inclusion chain \(\chainname{\examplegraphtwo} = \chain{\examplegraphtwo}{\chaindepthone}\) on \graphname{\examplegraphtwo}{} (of the same length as \chainname{\examplegraphone}) and a typing chain morphism \((\chainmorph{\examplegraphtwo},id_{[\chaindepthone]}) : \chainname{\examplegraphtwo} \to \chainname{\examplegraphone}\).

\section{Multilevel Typed Graph Transformations}
\label{sec:mcmt}


\subsubsection{Underlying Graph Transformation.}

To meet the characteristics of our application areas \cite{macias2016emisa,macias19jlamp,mantz2015coevolving} we work with cospans  
\begin{tikzpicture}[baseline=-1mm,node distance=13mm]
\node[el-math](l){\graphname{\rulegraphleft}{}};
\node[el-math](i)[right of=l]{\graphname{\rulegraphmiddle}{}};
\node[el-math](r)[right of=i]{\graphname{\rulegraphright}{}};
\draw[incmapl](l)to node[la-math,above](lm){\maprulelefttomiddle}(i);
\draw[incmapr](r)to	node[la-math,above](rm){\maprulerighttomiddle}(i);
\end{tikzpicture} 
of inclusion graph homomorphisms, where \(\rulegraphmiddle=\rulegraphleft\cup\rulegraphright\), as the \textbf{underlying graph transformation rule} of a multilevel typed rule. 
To apply such a rule~\cite{CorradiniHHK06,ehrig2006fundamentals,ehrig2009alternative}, we have to find a match \(\mapruletohierarchyleft:\graphname{\rulegraphleft}{}\to \graphname{\hierarchygraphleft}{}\) of \graphname{\rulegraphleft}{} in a graph \graphname{\hierarchygraphleft}{} at the bottom-most level of an MLM hierarchy. 
To describe the effect of a rule application, we adapt the Sq-PO approach \cite{CorradiniHHK06} to our 
cospan-rules:
First, we construct a pushout and, second, a final pullback complement (FPBC) to create the graphs  \graphname{\hierarchygraphmiddle}{} and \graphname{\hierarchygraphright}{}, resp. (see Fig.~\ref{fig:basic-constructions-rule-application}).
The details behind choosing cospan rules and Sq-PO, as opposed to span rules and double-pushout (DPO), are out of the scope of this paper. 
In short, however: 
(i) cospan rules are more suitable from an implementation point-of-view since they allow for first adding new elements then deleting (some of the) old elements~\cite{ehrig2009alternative}, and
(ii) having both old and new elements in $I$ allows us to introduce constraints on new elements depending on old constraints involving elements to be deleted~\cite{rutle2012formal}.
Moreover, we apply the rules using our variant of Sq-PO~\cite{CorradiniHHK06,ehrig2009alternative} since
(i) the pushout complement in DPO, even if it exists, may not be unique,
in contrast the FPBC, if it exists, is always unique (up to isomorphism),
(ii) FPBC allows faithful deletion in unknown context, i.e., dangling edges may be deleted by applying the rules, however, the co-match $\mapruletohierarchyright$ is always total, i.e., if the match $\mapruletohierarchyleft$ identifies elements to be removed with elements to be preserved, the FPBC will not exist and the application will not be allowed.

%
%

\vspace{-3ex}
\subsubsection{Multilevel Typed Rule.}
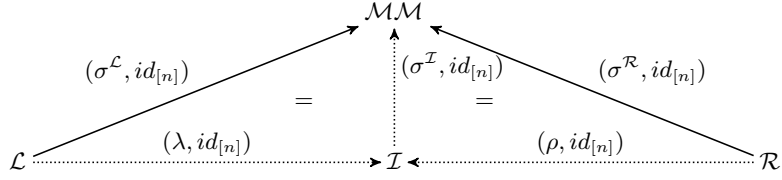
\begin{figure}[t]
	\begin{center}
		\begin{tikzpicture}[on grid,node distance=50mm]
		
		\node[el-math] (l0)							{};
		\node[el-math] (mm)	[right of=l0]						{\chainname{\rulegraph}};
		\node[el-math] (l)	[below=20mm of l0]		{\chainname{\rulegraphleft}};
		\node[el-math] (i)	[right of=l]		{\chainname{\rulegraphmiddle}};
		\node[el-math] (r)	[right of=i]		{\chainname{\rulegraphright}};
		\node[el-math] (e1)	[below left=17mm of mm]		{=};
		\node[el-math] (e2)	[below right=17mm of mm]	{=};
		
		\draw[map]	(l) to node[la-math]			(tl)	{(\chainmorph{\rulegraphleft},id_{[\chaindepthone]})}	(mm);
		\draw[map]	(r) to node[la-math,above right](tr)	{(\chainmorph{\rulegraphright},id_{[\chaindepthone]})}	(mm);
		\draw[mapdots]		(i) to node[la-math,above right]	(ti)	{(\chainmorph{\rulegraphmiddle},id_{[\chaindepthone]})}	(mm);
		
		\draw[mapdots]	(l) to node[la-math]			(lid)	{(\maprulelefttomiddle,id_{[\chaindepthone]})}			(i);
		\draw[mapdots]	(r) to node[la-math,above]			(rid)	{(\maprulerighttomiddle,id_{[\chaindepthone]})}			(i);
		\end{tikzpicture}
		\vspace{-1.5ex}
		\caption{Rule morphisms and their type compatibility}
		\label{fig:type-compatibility-rule-morphisms-by-typing-chain-morphisms}
	\end{center}
\vspace{-6mm}
\end{figure}

We augment the cospan rule to a \textbf{multilevel typed rule} by chosing a typing chain \(\chainname{\rulegraph}=(\tc{\rulegraph},\chaindepthone,\typemorph[\rulegraph]{}{})\), the typing chain of the rule, together with \textbf{coherent} multilevel typing's  over \(\chainname{\rulegraph}\) of \rulegraphleft\ and \rulegraphright, respectively.  
That is, we choose an inclusion chain \(\chainname{\rulegraphleft} = \chain{\rulegraphleft}{\chaindepthone}\) on \rulegraphleft, an inclusion chain \(\chainname{\rulegraphright} = \chain{\rulegraphright}{\chaindepthone}\) on \rulegraphright\ and typing chain morphisms 
\((\chainmorph{\rulegraphleft},id_{[\chaindepthone]}) : \chainname{\rulegraphleft} \to \chainname{\rulegraph}\) with \(\chainmorph{\rulegraphleft} =(\chainmorph[\indexone]{\rulegraphleft} :\graphname{\rulegraphleft}{\indexone} \to \graphname{\rulegraph}{\indexone} \mid \indexone \in [\chaindepthone])\),
\((\chainmorph{\rulegraphright},id_{[\chaindepthone]}) : \chainname{\rulegraphright} \to \chainname{\rulegraph}\) with \(\chainmorph{\rulegraphright} =(\chainmorph[\indexone]{\rulegraphright} :        \graphname{\rulegraphright}{\indexone} \to \graphname{\rulegraph}{\indexone} \mid \indexone \in [\chaindepthone])\) (see Fig.~\ref{fig:type-compatibility-rule-morphisms-by-typing-chain-morphisms}), such that \(\graphname{\rulegraphleft}{\indexone}\cap\graphname{\rulegraphright}{} =\graphname{\rulegraphleft}{}\cap\graphname{\rulegraphright}{\indexone}
=\graphname{\rulegraphleft}{\indexone}\cap\graphname{\rulegraphright}{\indexone}\) and, moreover,  \chainmorph[\indexone]{\rulegraphleft} and \chainmorph[\indexone]{\rulegraphright}
coincide on the intersection \(\graphname{\rulegraphleft}{\indexone}\cap\graphname{\rulegraphright}{\indexone}\) for all \(\indexone \in [\chaindepthone]\).

\begin{figure}[b]
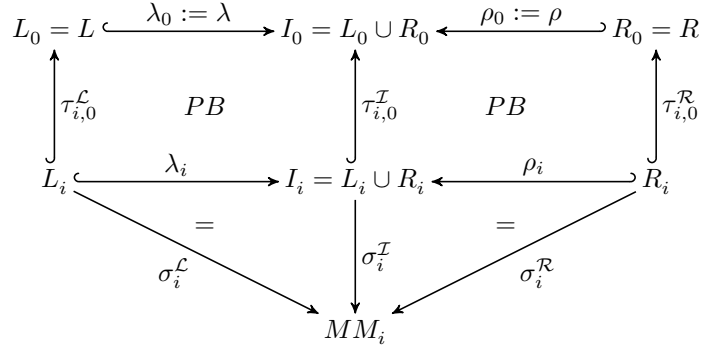

	\begin{center}
		\includestandalone{images/type-compatibility-rule-maps-3}
		\vspace{-2ex}
		\caption{Type compatibility of rule morphisms level-wise}
		\label{fig:type-compatibility-rule-morphisms-by-inclusion-chains}
	\end{center}
\end{figure}

The inclusion chain \(\chainname{\rulegraphmiddle} = \chain{\rulegraphmiddle}{\chaindepthone}\) on the union (pushout) \(\rulegraphmiddle=\rulegraphleft\cup\rulegraphright\)  is  simply constructed by level-wise unions (pushouts): \(\graphname{\rulegraphmiddle}{\indexone} := \graphname{\rulegraphleft}{\indexone}\cup\graphname{\rulegraphright}{\indexone}\)  for all \(\indexone \in [\chaindepthone]\); thus, we have \(\graphname{\rulegraphmiddle}{0}=\graphname{\rulegraphmiddle}{}\).
Since \cat{Graph} is an adhesive category \cite{ehrig2006fundamentals}, the construction of \chainname{\rulegraphmiddle} by pushouts and the coherence condition ensure that we get for any \(\indexone \in [\chaindepthone]\) two pullbacks as shown in Fig.~\ref{fig:type-compatibility-rule-morphisms-by-inclusion-chains}. 
The existence of these pullbacks  implies, according to the following Lemma, that we can reconstruct the inclusion chains \(\chainname{\rulegraphleft}\) and \(\chainname{\rulegraphright}\), respectively, as reducts of the inclusion chain \(\chainname{\rulegraphmiddle}\).

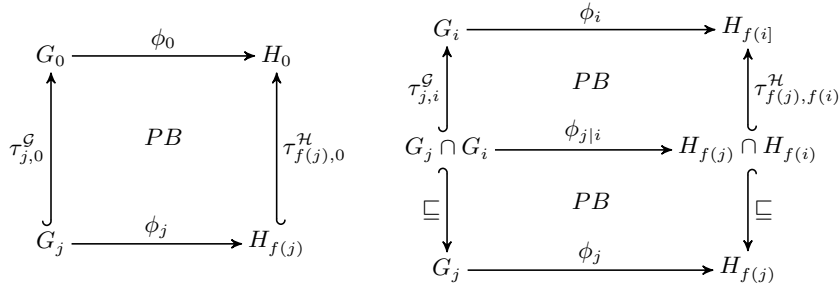
\begin{figure}[ht!]
	\begin{center}
		\begin{minipage}{.4\linewidth}
			\centering
			\begin{tikzpicture}[on grid,node distance=25mm]
			
			\node[el-math] (g0)						{\graphname{\examplegraphone}{0}};
			\node[el-math] (h0)	[right=30mm of g0]	{\graphname{\examplegraphtwo}{0}};
			\node[el-math] (gi)	[below of=g0]		{
			\graphname{\examplegraphone}{\indextwo}};
			\node[el-math] (hi)	[below of=h0]		{
			\graphname{\examplegraphtwo}{\maplevelone(\indextwo)}};
			
			\draw[map]		(gi) to node[la-math]		(pi)	{\morphone_\indextwo}										(hi);
			\draw[map]		(g0) to node[la-math]		(p0)	{\morphone_0}												(h0);
			\draw[incmapl]	(gi) to node[la-math]		(ing)	{\typemorph[\examplegraphone]{\indextwo}{0}}				(g0);
			\draw[incmapr]	(hi) to node[la-math,right]	(inh)	{\typemorph[\examplegraphtwo]{\maplevelone(\indextwo)}{0}}	(h0);
			
			\node[el-math]	(pb)	[below=13mm of p0]	{\pb};
			
			\end{tikzpicture}
		\end{minipage}
		\quad
		\begin{minipage}{.5\linewidth}
			\centering
			\begin{tikzpicture}[on grid,node distance=16mm]
			
			\node[el-math] (gi)						{\graphname{\examplegraphone}{\indexone}};
			\node[el-math] (hi)		[right=40mm of gi]{\graphname{\examplegraphtwo}{\maplevelone(\indexone]}};
			\node[el-math] (gji)	[below of=gi]	{
			\graphname{\examplegraphone}{\indextwo} \cap \graphname{\examplegraphone}{\indexone}};
			\node[el-math] (hji)	[below of=hi]	{
			\graphname{\examplegraphtwo}{\maplevelone(\indextwo)} \cap \graphname{\examplegraphtwo}{\maplevelone(\indexone)}};
			\node[el-math] (gj)		[below of=gji]	{\graphname{\examplegraphone}{\indextwo}};
			\node[el-math] (hj)		[below of=hji]	{\graphname{\examplegraphtwo}{\maplevelone(\indextwo)}};
			
			\draw[map]	(gi) to node[la-math]	(mgihi)		{\morphone_\indexone}					(hi);
			\draw[map]	(gj) to node[la-math]	(mgjhj)		{\morphone_\indextwo}					(hj);
			\draw[map]	(gji) to node[la-math]	(mgjihji)	{\morphone_{\indextwo \mid \indexone}}	(hji);
			
			\draw[incmapl]	(gji) to node[la-math]			(ingjigi)	{\typemorph[\examplegraphone]{\indextwo}{\indexone}}							(gi);
			\draw[incmapr]	(hji) to node[la-math,right]	(inhjihi)	{\typemorph[\examplegraphtwo]{\maplevelone(\indextwo)}{\maplevelone(\indexone)}}(hi);
			\draw[incmapr]	(gji) to node[la-math,left]		(ingjigj)	{\sqsubseteq}	(gj);
			\draw[incmapl]	(hji) to node[la-math,right]	(inhjihj)	{\sqsubseteq}	(hj);
			
			
			\node[el-math] (pb1)	[below=9mm of mgihi]	{\pb};
			\node[el-math] (pb2)	[below of=pb1]			{\pb};
			\end{tikzpicture}
		\end{minipage}
		\vspace{-2ex}
		\caption{Reduct of inclusion chains}
		\label{fig:reduction-of-inclusion chains}
	\end{center}
\vspace{-10mm}
\end{figure}

\begin{lemma}
	\label{lem:inclusion-chains-establish-chain-morphism}
	Let be given two inclusion chains \(\chainname{\examplegraphone} = \chain{\examplegraphone}{\chaindepthone}\) and \(\chainname{\examplegraphtwo} = \chain{\examplegraphtwo}{\chaindepthtwo}\) with \(\chaindepthone \leq \chaindepthtwo\) and a function \(\maplevelone : [\chaindepthone] \to [\chaindepthtwo]\) such that \(\maplevelone(0) = 0\) and  
	\(\indextwo > \indexone\) implies 
		\(\maplevelone(\indextwo) - \maplevelone(\indexone)\geq \indextwo - \indexone\)  for all \(\indexone, \indextwo \in [\chaindepthone]\).
    For any family \(\morphone = (\morphone_\indexone : \graphname{\examplegraphone}{\indexone} \to \graphname{\examplegraphtwo}{\maplevelone(\indexone)} \mid \indexone \in [\chaindepthone])\) of graph homomorphisms the following two requirements are equivalent:
	\begin{enumerate} 
		\item For all \(\chaindepthone \geq \indextwo > 0\) the left-hand square in Fig.~\ref{fig:reduction-of-inclusion chains} is a pullback.
		\item The pair \((\morphone,\maplevelone)\) constitutes a closed typing chain morphism \((\morphone,\maplevelone) : \chainname{\examplegraphone} \to \chainname{\examplegraphtwo}\) where for all  \(\chaindepthone \geq \indextwo >\indexone \geq 0\) the right-hand diagram in Fig.~\ref{fig:reduction-of-inclusion chains} consists of two pullbacks.
	\end{enumerate}
\end{lemma}

Given a closed typing chain morphism \((\morphone,\maplevelone) : \chainname{\examplegraphone} \to \chainname{\examplegraphtwo}\) between inclusion chains, as described in Lemma~\ref{lem:inclusion-chains-establish-chain-morphism}, we call \(\chainname{\examplegraphone}\) the \textbf{reduct of \(\chainname{\examplegraphtwo}\) along \(\morphone_0 : \graphname{\examplegraphone}{0} \to \graphname{\examplegraphtwo}{0}\) and \(\maplevelone : [\chaindepthone] \to [\chaindepthtwo]\) } while \((\morphone,\maplevelone) : \chainname{\examplegraphone} \to \chainname{\examplegraphtwo}\) is called a \textbf{reduct morphism}.
Note that the composition of two reduct morphisms is a reduct morphism as well.

Lemma \ref{lem:inclusion-chains-establish-chain-morphism} ensures that the families 
 \((\maprulelefttomiddle_{\indexone}:\graphname{\rulegraphleft}{\indexone} \to \graphname{\rulegraphmiddle}{\indexone} \mid \indexone \in [\chaindepthone])\) and \((\maprulerighttomiddle_{\indexone}:\graphname{\rulegraphright}{\indexone} \to \graphname{\rulegraphmiddle}{\indexone} \mid \indexone \in [\chaindepthone])\) of inclusion graph homomorphisms establish reduct morphisms \((\maprulelefttomiddle,id_{[\chaindepthone]}): \chainname{\rulegraphleft}\to \chainname{\rulegraphmiddle}\) and 
\((\maprulerighttomiddle,id_{[\chaindepthone]}): \chainname{\rulegraphright}\to \chainname{\rulegraphmiddle}\), resp., as shown in Fig.~\ref{fig:type-compatibility-rule-morphisms-by-typing-chain-morphisms}.

Finally, we have to construct a typing chain morphism \((\chainmorph{\rulegraphmiddle},id_{[\chaindepthone]}) : \chainname{\rulegraphmiddle} \to \chainname{\rulegraph}\) 
making the diagram 
in Fig.~\ref{fig:type-compatibility-rule-morphisms-by-typing-chain-morphisms} commute:  For all \(\indexone \in [\chaindepthone]\), we constructed the union (pushout)  \(\graphname{\rulegraphmiddle}{\indexone} := \graphname{\rulegraphleft}{\indexone}\cup\graphname{\rulegraphright}{\indexone}\). Moreover, \chainmorph[\indexone]{\rulegraphleft} and \chainmorph[\indexone]{\rulegraphright}
coincide on \(\graphname{\rulegraphleft}{\indexone}\cap\graphname{\rulegraphright}{\indexone}\), by coherence assumption, thus we get a unique  \(\chainmorph[\indexone]{\rulegraphmiddle} :\graphname{\rulegraphmiddle}{\indexone} \to \graphname{\rulegraph}{\indexone}\) such that (see Fig.~\ref{fig:type-compatibility-rule-morphisms-by-inclusion-chains}) 
\begin{equation}\label{eq:type-compatibility-rule-morphisms-by-inclusion-chains}
\chainmorph[\indexone]{\rulegraphleft} = \maprulelefttomiddle_{\indexone};\chainmorph[\indexone]{\rulegraphmiddle} 
\quad\mbox{and}\quad 
\chainmorph[\indexone]{\rulegraphright} = \maprulerighttomiddle_{\indexone};\chainmorph[\indexone]{\rulegraphmiddle} 
\end{equation}
Since \cat{Graph} is adhesive, Lemma \ref{lem:inclusion-chains-establish-chain-morphism}
ensures that the family \(\chainmorph{\rulegraphmiddle} =(\chainmorph[\indexone]{\rulegraphmiddle} :        \graphname{\rulegraphmiddle}{\indexone} \to \graphname{\rulegraph}{\indexone} \mid \indexone \in [\chaindepthone])\) of graph homomorphisms establishes indeed a typing chain morphism \((\chainmorph{\rulegraphmiddle},id_{[\chaindepthone]}) : \chainname{\rulegraphmiddle} \to \chainname{\rulegraph}\) while the equations \myref{eq:type-compatibility-rule-morphisms-by-inclusion-chains} ensure that the diagram in Fig.~\ref{fig:type-compatibility-rule-morphisms-by-typing-chain-morphisms} commutes indeed.

\begin{wrapfigure}[10]{l}{.5\textwidth}
  \centering
  \includestandalone[scale=.8]{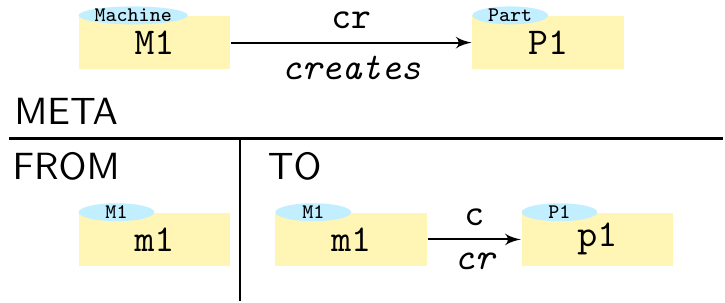}
	\caption{\textit{CreatePart}: a sample rule}
	\label{fig:pls-rule-create-part}
\end{wrapfigure}

\begin{example}
\label{ex:multilevel-typed-rule}
Fig.~\ref{fig:pls-rule-create-part} shows a multilevel typed rule \textit{CreatePart} from a case study \cite{macias19jlamp}.  This rule can be used to specify the behaviour of machines that create parts, by matching an existing type of machine that generates a certain type of parts, and in the instance at the bottom, generating such a part.
\textsf{META} defines a typing chain \chainname{\rulegraph} of depth \(3\).
It declares the graph 
(\begin{tikzpicture}[baseline=-1mm,node distance=12mm]
\node[el-math](m1){\elementname{M1}};
\node[el-math](p1)[right of=m1]{\elementname{P1}};
\draw[map](m1)to	node[la-math,above](rm){\elementname{cr}}(p1);
\end{tikzpicture})
that becomes \graphname{\rulegraph}{2}. The declaration of the direct types \elementname{Machine}, \elementname{creates}, \elementname{Part} for the elements in \graphname{\rulegraph}{2} declares, implicitly, a graph \(\graphname{\rulegraph}{1}:=\)
(\begin{tikzpicture}[baseline=-1mm,node distance=28mm]
\node[el-math](m1){\elementname{Machine}};
\node[el-math](p1)[right of=m1]{\elementname{Part}};
\draw[map](m1)to	node[la-math,above](rm){\elementname{creates}}(p1);
\end{tikzpicture})
that is in turn, implicitly, typed over \(\graphname{\rulegraph}{0}:=\elementname{ECore}\). All the  morphisms in \typemorph[\rulegraph]{}{} are total and uniquely determined thus we have, especially,  \(\typemorph[\rulegraph]{2}{0}=\typemorph[\rulegraph]{2}{1};\typemorph[\rulegraph]{1}{0}\).

\textsf{FROM} and \textsf{TO} declare as well the left-hand side \(\rulegraphleft:=(\,\elementname{m1}\,)\) and the right-hand-side \(\rulegraphright:=\) 
(\begin{tikzpicture}[baseline=-1mm,node distance=10mm]
\node[el-math](m1){\elementname{m1}};
\node[el-math](p1)[right of=m1]{\elementname{p1}};
\draw[map](m1)to	node[la-math,above](rm){\elementname{c}}(p1);
\end{tikzpicture}),
resp., of the rule and the direct types of the elements in  \rulegraphleft\ and \rulegraphright. These direct types are all located in \graphname{\rulegraph}{2} thus \(\graphname{L}{2}=\graphname{L}{}\) and \(\graphname{R}{2}=\graphname{R}{}\) where the direct types define nothing but the typing morphisms \(\chainmorph[2]{\rulegraphleft}:\graphname{\rulegraphleft}{2}\to\graphname{\rulegraph}{2}\) and \(\chainmorph[2]{\rulegraphright}:\graphname{\rulegraphright}{2}\to\graphname{\rulegraph}{2}\), resp.
The other typing morphisms are obtained by ``transitive closure'', i.e., \(\chainmorph[1]{\rulegraphleft}:=\chainmorph[2]{\rulegraphleft};\typemorph[\rulegraph]{2}{1}\),  \(\chainmorph[0]{\rulegraphleft}:=\chainmorph[2]{\rulegraphleft};\typemorph[\rulegraph]{2}{0}\) and  \(\chainmorph[1]{\rulegraphright}:=\chainmorph[2]{\rulegraphright};\typemorph[\rulegraph]{2}{1}\),  \(\chainmorph[0]{\rulegraphright}:=\chainmorph[2]{\rulegraphright};\typemorph[\rulegraph]{2}{0}\),
thus we have
\(\graphname{L}{}=\graphname{L}{0}=\graphname{L}{1}=\graphname{L}{2}\) and \(\graphname{R}{}=\graphname{R}{0}=\graphname{R}{1}=\graphname{R}{2}\).

For the ``plain variant'' of the rule \textit{CreatePart} (in Fig.~\ref{fig:create-part-modified-and-application}), \chainname{\rulegraph} consists only of the graphs \graphname{\rulegraph}{1} =
(\begin{tikzpicture}[baseline=-1mm,node distance=12mm]
\node[el-math](m1){\elementname{M1}};
\node[el-math](p1)[right of=m1]{\elementname{P1}};
\draw[map](m1)to	node[la-math,above](rm){\elementname{cr}}(p1);
\end{tikzpicture}),
 \(\graphname{\rulegraph}{0}=\elementname{ECore}\) 
 and the trivial \(\typemorph[\rulegraph]{1}{0}\).
\end{example}
\subsubsection{Multilevel Typed Match.}

In the multilevel typed setting all the graphs \graphname{\hierarchygraphleft}{}, \graphname{\hierarchygraphmiddle}{}, \graphname{\hierarchygraphright}{} are multilevel typed over a common typing chain \(\chainname{\hierarchygraph}=(\tc{\hierarchygraph},\chaindepthtwo,\typemorph[\hierarchygraph]{}{})\), with \(\chaindepthone\leq\chaindepthtwo\),  that is determined by the path from \graphname{\hierarchygraphleft}{} to the top of the current MLM hierarchy (see Fig.~\ref{fig:mlm-scheme}).

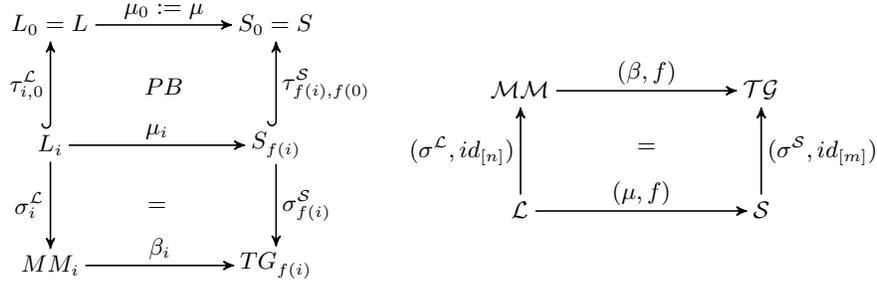
\begin{figure}[ht]
	\begin{center}
		\begin{minipage}{.4\linewidth}
			\centering
		\begin{tikzpicture}[on grid,node distance=16mm]
		
		\node[el-math] (l0)     				{\graphname{\rulegraphleft}{0} = \graphname{\rulegraphleft}{}};
		\node[el-math] (s0) [right=30mm of l0]		{\graphname{\hierarchygraphleft}{0} = \graphname{\hierarchygraphleft}{}};
		\node[el-math] (li)	[below of=l0]		{\graphname{\rulegraphleft}{\indexone}};
		\node[el-math] (si)	[below of=s0]		{\graphname{\hierarchygraphleft}{\maplevelone(\indexone)}};
		\node[el-math] (mm)	[below of=li]					{\graphname{\rulegraph}{\indexone}};
		\node[el-math] (tg)	[below of=si]	{\graphname{\hierarchygraph}{\maplevelone(\indexone)}};
		
		\draw[map]		(mm) to node[la-math]		(bi)	{\mapruletohierarchybinding_{\indexone}}									(tg);
		\draw[map]		(l0) to node[la-math]		(m0)	{\mapruletohierarchyleft_{0} := \mapruletohierarchyleft}						(s0);
		\draw[map]		(li) to node[la-math]		(mi)	{\mapruletohierarchyleft_{\indexone}}										(si);
		\draw[incmapl]	(li) to node[la-math,left]	(tl)	{\typemorph[\rulegraphleft]{\indexone}{0}}									(l0);
		\draw[incmapl]	(si) to node[la-math,right]	(ts)	{\typemorph[\hierarchygraphleft]{\maplevelone(\indexone)}{\maplevelone(0)}}	(s0);
		%
		\draw[map]		(li) to node[la-math,left]	(sl2)	{\chainmorph[\indexone]{\rulegraphleft}}					(mm);
		\draw[map]		(si) to node[la-math,right]	(ss2)	{\chainmorph[\maplevelone(\indexone)]{\hierarchygraphleft}}	(tg);
		\node[el-math] (eq)	[below=10mm of mi]	{=};
		\node[el-math] (pb) [below=10mm of m0]		{\pb};
		
		\end{tikzpicture}
		\end{minipage}
		\quad
		\begin{minipage}{.5\linewidth}
			\centering
		\begin{tikzpicture}[on grid,node distance=32mm]
		
		\node[el-math] (mm)						{\chainname{\rulegraph}};
		\node[el-math] (tg)	[right of=mm]		{\chainname{\hierarchygraph}};
		\node[el-math] (l)	[below=16mm of mm]	{\chainname{\rulegraphleft}};
		\node[el-math] (s)	[right of=l]		{\chainname{\hierarchygraphleft}};
		
		\draw[map]	(l)	 to node[la-math]		(lmm)	{(\chainmorph{\rulegraphleft},id_{[\chaindepthone]})}		(mm);
		\draw[map]	(mm) to node[la-math]		(mmtg)	{(\mapruletohierarchybinding,\maplevelone)}					(tg);
		\draw[map]	(s)	 to node[la-math,right]	(stg)	{(\chainmorph{\hierarchygraphleft},id_{[\chaindepthtwo]})}	(tg);		
		\draw[map]	(l) to node[la-math]		(lid)	{(\mapruletohierarchyleft,\maplevelone)}					(s);
		
		\node[el-math] (e)	[below=10mm of mmtg]{=};
		
		\end{tikzpicture}
		\end{minipage}
		
		\caption{Conditions for Multilevel Typed Match}
		\vspace{-1.5ex}
		\label{fig:conditions-multilevel-typed-match}
	\end{center}
\vspace{-6mm}
\end{figure}

A \textbf{match} of the multilevel typed rule into a graph \hierarchygraphleft\ with a given multilevel typing over \chainname{\hierarchygraph}, i.e., an inclusion chain \(\chainname{\hierarchygraphleft} = \chain{\hierarchygraphleft}{\chaindepthtwo}\)  with \(\graphname{\hierarchygraphleft}{0}=\graphname{\hierarchygraphleft}{}\) and  a typing chain morphism \((\chainmorph{\hierarchygraphleft},id_{[\chaindepthtwo]}) : \chainname{\hierarchygraphleft} \to \chainname{\hierarchygraph}\),
is given by a graph homomorphism \(\mapruletohierarchyleft:\rulegraphleft\to\hierarchygraphleft\) and a typing chain morphism \((\mapruletohierarchybinding,\maplevelone) : \chainname{\rulegraph} \to \chainname{\hierarchygraph}\) 
such that the following two conditions are satisfied:
\begin{itemize}
    \item \textbf{Reduct:} \chainname{\rulegraphleft} is the reduct of \chainname{\hierarchygraphleft} along \(\mapruletohierarchyleft:\rulegraphleft\to\hierarchygraphleft\) and \(\maplevelone:[\chaindepthone]\to[\chaindepthtwo]\), i.e., \(\mapruletohierarchyleft_{0} := \mapruletohierarchyleft: \graphname{\rulegraphleft}{0} = \graphname{\rulegraphleft}{} \longrightarrow \graphname{\hierarchygraphleft}{0} = \graphname{\hierarchygraphleft}{}\) extends uniquely (by pullbacks) to a reduction morphism \((\mapruletohierarchyleft,\maplevelone): \chainname{\rulegraphleft}\to \chainname{\hierarchygraphleft}\) with \(\mapruletohierarchyleft= (\mapruletohierarchyleft_{\indexone}:\graphname{\rulegraphleft}{\indexone} \to \graphname{\hierarchygraphleft}{\maplevelone(\indexone)} \mid \indexone \in [\chaindepthone])\) (see Fig.~\ref{fig:conditions-multilevel-typed-match}).
    \item \textbf{Type compatibility:} \((\chainmorph{\rulegraphleft},id_{[\chaindepthone]});(\mapruletohierarchybinding,\maplevelone) = (\mapruletohierarchyleft,\maplevelone);(\chainmorph{\hierarchygraphleft},id_{[\chaindepthtwo]})\), i.e., we require 
    \vspace{-1ex}
\begin{equation}
\label{eq:type-compatibility-rule-match-by-inclusion-chains}
\chainmorph[\indexone]{\rulegraphleft};\mapruletohierarchybinding_{\indexone} = \mapruletohierarchyleft_{\indexone};\chainmorph[\maplevelone(\indexone)]{\hierarchygraphleft} \mbox{ for all } \chaindepthone\geq\indexone > 0.
\end{equation}
\end{itemize}
\vspace{-3ex}

\vspace{-6mm}
\subsubsection{Application of a Multilevel Typed Rule -- Objectives.} The basic idea is to construct for a given application of a graph transformation rule, as shown in Fig.~\ref{fig:basic-constructions-rule-application}, a unique type compatible multilevel typing of the result graphs \hierarchygraphmiddle\ and \hierarchygraphright.
The parameters of this construction are typing chains \chainname{\rulegraph}, \chainname{\hierarchygraph}; a coherent multilevel typing of the graph transformation rule over \chainname{\rulegraph}; a multilevel typing of the graph \hierarchygraphleft\ over \chainname{\hierarchygraph} and a typing chain morphism \((\mapruletohierarchybinding,\maplevelone) : \chainname{\rulegraph} \to \chainname{\hierarchygraph}\) extending the given match \(\mapruletohierarchyleft:\rulegraphleft\to\hierarchygraphleft\) of graphs to a multilevel typed match satisfying the two respective conditions for multilevel typed matches.

\begin{wrapfigure}[13]{l}{.59\textwidth}
\vspace{-5ex}
\begin{center}
	\begin{tikzpicture}[on grid,node distance=27mm]
	
	\node[el-math] (l)										{\chainname{\rulegraphleft}};
	\node[el-math] (i)	[right of=l]						{\chainname{\rulegraphmiddle}};
	\node[el-math] (mm)	[above of=i]						{\chainname{\rulegraph}};
	\node[el-math] (s)	[below right=10mm and 25mm of l]	{\chainname{\hierarchygraphleft}};
	\node[el-math] (d)	[right of=s]						{\chainname{\hierarchygraphmiddle}};
	\node[el-math] (tg)	[above of=d]						{\chainname{\hierarchygraph}};
	
	\draw[map]		(l) to node[la-math,below left]		(mf)	{(\mapruletohierarchyleft,\maplevelone)}						(s);
	\draw[incmapl] 		(l) to node[la-math]				(li)	{(\maprulelefttomiddle,id_{[\chaindepthone]})}					(i);
	\draw[incmapl] 	(s) to node[la-math,pos=.35]			(sd)	{(\maphierarchylefttomiddle,id_{[\chaindepthtwo]})}				(d);
	\draw[mapdots]	(d) to node[la-math,right,pos=.3]	(dt)	{(\chainmorph{\hierarchygraphmiddle},id_{[\chaindepthtwo]})}	(tg);
	\draw[map]		(i) to node[la-math,pos=.3]			(im)	{(\chainmorph{\rulegraphmiddle},id_{[\chaindepthone]})}			(mm);
	\draw[map]		(s) to node[la-math,pos=.85]			(st)	{(\chainmorph{\hierarchygraphleft},id_{[\chaindepthtwo]})}		(tg);
	\draw[map]		(l) to node[la-math]				(lm)	{(\chainmorph{\rulegraphleft},id_{[\chaindepthone]})}			(mm);
	\draw[map]		(i) to node[la-math]				(df)	{(\mapruletohierarchymiddle,\maplevelone)}						(d);
	\draw[map]		(mm) to	node[la-math]				(bf)	{(\mapruletohierarchybinding,\maplevelone)}						(tg);
	
	\end{tikzpicture}
\end{center}
	\vspace{-4ex}
	\caption{Pushout step}
	\label{fig:pushout-step}
\end{wrapfigure}
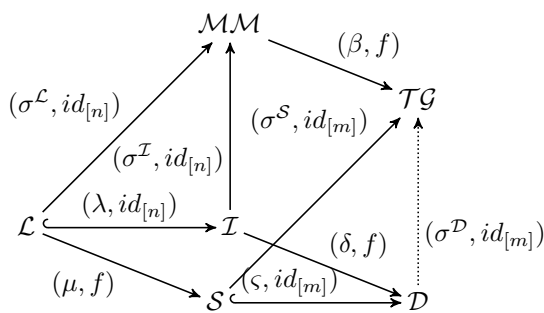

\begin{example}[Multilevel Typed Match]
\label{ex:match+rule-application}
To achieve precision in rule application the elements \elementname{Machine}, \elementname{creates}, \elementname{Part} in the original rule \textit{CreatePart} are constants required to match syntactically with elements in the hierarchy. In such a way, \graphname{\rulegraph}{1} = 
(\begin{tikzpicture}[baseline=-1mm,node distance=28mm]
\node[el-math](m1){\elementname{Machine}};
\node[el-math](p1)[right of=m1]{\elementname{Part}};
\draw[map](m1)to	node[la-math,above](rm){\elementname{creates}}(p1);
\end{tikzpicture})
has to match with \elementname{generic\_plant} while \graphname{\rulegraph}{2} =
(\begin{tikzpicture}[baseline=-1mm,node distance=12mm]
\node[el-math](m1){\elementname{M1}};
\node[el-math](p1)[right of=m1]{\elementname{P1}};
\draw[map](m1) to	node[la-math,above](rm){\elementname{cr}}(p1);
\end{tikzpicture})
could match with \elementname{hammer\_plant} or \elementname{stool\_plant}.
We will observe later that for the plain version of the rule \textit{CreatePart} in Fig.~\ref{fig:create-part-modified-and-application} we could match \graphname{\rulegraph}{1} =
(\begin{tikzpicture}[baseline=-1mm,node distance=12mm]
\node[el-math](m1){\elementname{M1}};
\node[el-math](p1)[right of=m1]{\elementname{P1}};
\draw[map](m1) to	node[la-math,above](rm){\elementname{cr}}(p1);
\end{tikzpicture}) 
either with \graphname{\hierarchygraph}{2} = \elementname{hammer\_plant} or \graphname{\hierarchygraph}{1} = \elementname{generic\_plant} in the hierarchy in Fig.~\ref{fig:hierarchy-introductory-example-4},  where the second match would lead to undesired results (see Example \ref{ex:rule-application}).
%
\end{example}
\noindent
\textbf{Pushout step.}
As shown later, the pushout of the span
\begin{tikzpicture}[baseline=-1mm,node distance=15mm]
\node[el-math](s){\graphname{\hierarchygraphleft}{}};
\node[el-math](l)[right of=s]{\graphname{\rulegraphleft}{}};
\node[el-math](i)[right of=l]{\graphname{\rulegraphmiddle}{}};
\draw[incmapl](l)to node[la-math,above](lm){\maprulelefttomiddle}(i);
\draw[map](l)to	node[la-math,above](rm){\mapruletohierarchyleft}(s);
\end{tikzpicture} 
in \cat{Graph} extends, in a canonical way, to a pushout of the span 
\begin{center}
\vspace{-1ex}
	\begin{tikzpicture}[on grid,node distance=35mm]
	
	\node[el-math] (s)					{\chainname{\hierarchygraphleft}};
	\node[el-math] (l)	[right of=s]	{\chainname{\rulegraphleft}};
	\node[el-math] (i)	[right of=l]	{\chainname{\rulegraphmiddle}};
	
	\draw[map]		(l) to node[la-math,above]	(map)	{(\mapruletohierarchyleft,\maplevelone)}		(s);
	\draw[incmapl]	(l) to node[la-math,above]	(in)	{(\maprulelefttomiddle,id_{[\chaindepthone]})}	(i);
	\end{tikzpicture}
	\vspace{-1ex}
\end{center}
of reduct morphisms in \cat{Chain} such that the result typing chain \(\chainname{\hierarchygraphmiddle} = \chain{\hierarchygraphmiddle}{\chaindepthtwo}\) is an inclusion chain and the typing chain morphisms 
\((\maphierarchylefttomiddle, id_{[\chaindepthtwo]}) : \chainname{\hierarchygraphleft} \hookrightarrow \chainname{\hierarchygraphmiddle}\) and  \((\mapruletohierarchymiddle,\maplevelone):\chainname{\rulegraphmiddle}\to \chainname{\hierarchygraphmiddle}\) become reduct morphisms (see the bottom in Fig.~\ref{fig:pushout-step}).

We get also a type compatible typing chain morphism from \chainname{\hierarchygraphmiddle} into \chainname{\hierarchygraph}:
The back triangle in Fig.~\ref{fig:pushout-step} commutes due to the type compatibility of the rule (see Fig.~\ref{fig:type-compatibility-rule-morphisms-by-typing-chain-morphisms}). The roof square commutes since the match 
is type compatible (see Fig.~\ref{fig:conditions-multilevel-typed-match}).
This gives us \((\mapruletohierarchyleft , \maplevelone) ; (\chainmorph{\hierarchygraphleft},id_{[\chaindepthtwo]}) = (\maprulelefttomiddle , id_{[\chaindepthone]});(\chainmorph{\rulegraphmiddle} , id_{[\chaindepthone]}) ; (\mapruletohierarchybinding , \maplevelone)\), thus the universal property of the pushout bottom square provides a unique chain morphism \((\chainmorph{\hierarchygraphmiddle} , id_{[\chaindepthtwo]}) : \chainname{\hierarchygraphmiddle} \to \chainname{\hierarchygraph}\) such that both type compatibility conditions \((\maphierarchylefttomiddle , id_{[\chaindepthtwo]}) ; (\chainmorph{\hierarchygraphmiddle} , id_{[\chaindepthtwo]}) = (\chainmorph{\hierarchygraphleft} , id_{[\chaindepthtwo]})\) and \((\mapruletohierarchymiddle , \maplevelone);(\chainmorph{\hierarchygraphmiddle} , id_{[\chaindepthtwo]}) = (\chainmorph{\rulegraphmiddle} , id_{[\chaindepthone]}) ; (\mapruletohierarchybinding , \maplevelone)\) are satisfied.

\noindent
\subsubsection{Pullback complement step.} 
\begin{figure}[ht!]
	\begin{center}
		\begin{tikzpicture}[on grid,node distance=30mm]
		
		\node[el-math] (i)									{\chainname{\rulegraphmiddle}};
		\node[el-math] (r)	[right of=i]					{\chainname{\rulegraphright}};
		\node[el-math] (mm)	[above of=i]					{\chainname{\rulegraph}};
		\node[el-math] (d)	[below right=10mm and 25mm of i]{\chainname{\hierarchygraphmiddle}};
		\node[el-math] (t)	[right of=d]					{\chainname{\hierarchygraphright}};
		\node[el-math] (tg)	[above of=d]					{\chainname{\hierarchygraph}};
		
		\draw[mapdots]		(r) to node[la-math,below left,pos=.3]	(mf)	{(\mapruletohierarchyright,\maplevelone)}						(t);
		\draw[map] 		(r) to node[la-math,below,pos=.4]		(li)	{(\maprulerighttomiddle,id_{[\chaindepthone]})}					(i);
		\draw[mapdots] 		(t) to node[la-math,below]				(sd)	{(\maphierarchyrighttomiddle,id_{[\chaindepthtwo]})}			(d);
		\draw[map]		(d) to node[la-math,right]				(dt)	{(\chainmorph{\hierarchygraphmiddle},id_{[\chaindepthtwo]})}	(tg);
		\draw[map]		(i) to node[la-math]					(im)	{(\chainmorph{\rulegraphmiddle},id_{[\chaindepthone]})}			(mm);
		\draw[mapdots]	(t) to node[la-math,above right,pos=.7]	(st)	{(\chainmorph{\hierarchygraphright},id_{[\chaindepthtwo]})}		(tg);
		\draw[map]		(r) to node[la-math,pos=.4]				(lm)	{(\chainmorph{\rulegraphright},id_{[\chaindepthone]})}			(mm);
		\draw[map]		(i) to node[la-math,below left]			(df)	{(\mapruletohierarchymiddle,\maplevelone)}						(d);
		\draw[map]		(mm) to	node[la-math]					(bf)	{(\mapruletohierarchybinding,\maplevelone)}						(tg);
		
		\end{tikzpicture}
	\end{center}
	\vspace{-4ex}
	\caption{Pullback complement step}
	\label{fig:pullback-complement-step}
	\vspace{-3ex}
\end{figure}
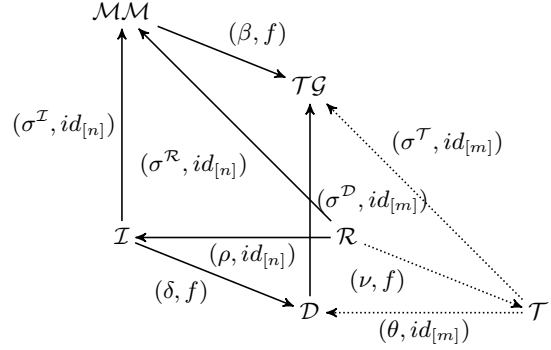

As shown later, the final pullback complement 
\begin{tikzpicture}[baseline=-1mm,node distance=15mm]
\node[el-math](d){\graphname{\hierarchygraphmiddle}{}};
\node[el-math](t)[right of=d]{\graphname{\hierarchygraphright}{}};
\node[el-math](r)[right of=t]{\graphname{\rulegraphright}{}};
\draw[incmapr](r)to node[la-math,above](lm){\mapruletohierarchyright}(t);
\draw[map](t)to	node[la-math,above](rm){\maphierarchyrighttomiddle}(d);
\end{tikzpicture} 
in \cat{Graph}
extends, in a canonical way, to  a sequence of reduct morphisms
	\begin{tikzpicture}[baseline=-1mm,node distance=25mm]
	
	\node[el-math] (d)					{\chainname{\hierarchygraphmiddle}};
	\node[el-math] (t)	[right of=d]	{\chainname{\hierarchygraphright}};
	\node[el-math] (r)	[right of=t]	{\chainname{\rulegraphright}};
	
	\draw[map]		(r) to node[la-math,above]	(map)	{(\mapruletohierarchyright,\maplevelone)}		(t);
	\draw[incmapr]	(t) to node[la-math,above]	(in)	{(\maphierarchyrighttomiddle,id_{[\chaindepthone]})}	(d);
	
	\end{tikzpicture}
in \cat{Chain} such that the bottom square in Fig.~\ref{fig:pullback-complement-step} commutes.
%

\vspace{-4ex}


\subsubsection{Pushout of reduct morphisms -- Two steps.}
\label{subsec:pushouts-category-chain}
\begin{figure}[b]
	\begin{center}
		\begin{tikzpicture}[on grid,node distance=18mm]
		
		\node[el-math] (l)										{\chainname{\rulegraphleft}};
		\node[el-math] (sf)	[right=40mm of l]					{\chainname[\downarrow\maplevelone]{\hierarchygraphleft}};
		\node[el-math] (s)	[right=40mm of sf]					{\chainname{\hierarchygraphleft}};
		\node[el-math] (i)	[below of=l]						{\chainname{\rulegraphmiddle}};
		\node[el-math] (df)	[below of=sf]						{\chainname[\downarrow\maplevelone]{\hierarchygraphmiddle}};
		\node[el-math] (d)	[below of=s]						{\chainname{\hierarchygraphmiddle}};
		\node[el-math] (eq)	[below right=9mm and 20mm of l]	{(1)};
		\node[el-math] (eq)	[below right=9mm and 25mm of sf]	{(2)};
		
		\draw[map]		(l) to node[la-math]			(mu1)	{(\mapruletohierarchyleft,id_{[\chaindepthone]})}											(sf);
		\draw[map,bend left](l) to node[la-math,below]		(muf)	{(\mapruletohierarchyleft,\maplevelone)}													(s);
		\draw[map]		(sf) to node[la-math]			(id1)	{(\overline{id}^{\chainname{\hierarchygraphleft}}_{\downarrow\maplevelone},\maplevelone)}	(s);
		\draw[incmapl]	(l) to node[la-math,left]		(in1)	{(\maprulelefttomiddle,id_{[\chaindepthone]})}												(i);
		\draw[incmapl]	(sf) to node[la-math]			(in2)	{(\maphierarchylefttomiddle_{\downarrow\maplevelone},id_{[\chaindepthone]})}				(df);
		\draw[incmapl]	(s) to node[la-math]			(in3)	{(\maphierarchylefttomiddle,id_{[\chaindepthtwo]})}											(d);
		\draw[map]		(i) to node[la-math,below]		(mu2)	{(\mapruletohierarchymiddle,id_{[\chaindepthone]})}											(df);
		\draw[map,bend right](i) to node[la-math]		(def)	{(\mapruletohierarchymiddle,\maplevelone)}													(d);
		\draw[map]	(df) to node[la-math,below]		(id2)	{(\overline{id}^{\chainname{\hierarchygraphmiddle}}_{\downarrow\maplevelone},\maplevelone)}	(d);
		
		\end{tikzpicture}
	\end{center}
	\vspace{-5ex}
	\caption{Two pushout steps to construct the inclusion chain \(\chainname{\hierarchygraphmiddle}\)}
	\label{fig:two-po-construct-d}
\end{figure}
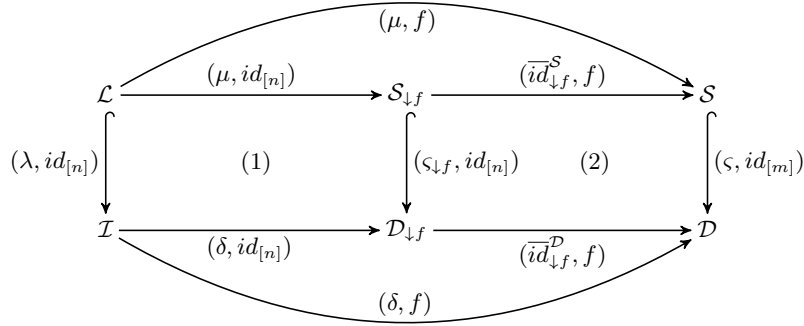

We discuss the intended pushout of the  span
\vspace{-2.5ex}
\begin{center}
	\begin{tikzpicture}[on grid,node distance=35mm]
	
	\node[el-math] (s)					{\chainname{\hierarchygraphleft}};
	\node[el-math] (l)	[right of=s]	{\chainname{\rulegraphleft}};
	\node[el-math] (i)	[right of=l]	{\chainname{\rulegraphmiddle}};
	
	\draw[map]		(l) to node[la-math,above]	(map)	{(\mapruletohierarchyleft,\maplevelone)}		(s);
	\draw[incmapl]	(l) to node[la-math,above]	(in)	{(\maprulelefttomiddle,id_{[\chaindepthone]})}	(i);
	
	\end{tikzpicture}
\vspace{-1.5ex}
\end{center}
of reduct morphisms in \(\cat{Chain}\). 
%
%
%
%
The reduct morphism \((\maprulelefttomiddle,id_{[\chaindepthone]})\) is surjective w.r.t.\ levels, thus the pushout inclusion chain \chainname{\hierarchygraphmiddle} should have the same length as \chainname{\hierarchygraphleft}.
%
The rule provides, however, only information how to extend the subgraphs of \(\graphname{\hierarchygraphleft}{0} = \graphname{\hierarchygraphleft}{}\) at the levels \(\maplevelone([\chaindepthone]) \subseteq [\chaindepthtwo]\).
For the subgraphs in \(\chainname{\hierarchygraphleft}\) at levels in \([\chaindepthtwo] \setminus \maplevelone([\chaindepthone])\) the rule does not impose anything thus we let the subgraphs at those levels untouched.
In terms of typing chain morphisms, this means that we factorize the reduct morphism \((\mapruletohierarchyleft,\maplevelone)\) into two reduct morphisms and that we will construct the resulting inclusion chain \chainname{\hierarchygraphmiddle} in two pushout steps (see Fig.~\ref{fig:two-po-construct-d}) where \(\chainname[\downarrow\maplevelone]{\hierarchygraphleft} := \chain[\downarrow\maplevelone]{\hierarchygraphleft}{\chaindepthone}\) with
\(\overline{\hierarchygraphleft}_{\downarrow \maplevelone} := [\graphname{\hierarchygraphleft}{\maplevelone(\chaindepthone)},  \graphname{\hierarchygraphleft}{\maplevelone(\chaindepthone-1)}, \ldots, \graphname{\hierarchygraphleft}{\maplevelone(1)}, \graphname{\hierarchygraphleft}{\maplevelone(0)=0}]\)
and
\(\typemorph[\hierarchygraphleft]{}{}_{\downarrow\maplevelone} := (\typemorph[\hierarchygraphleft]{\maplevelone(\indextwo)}{\maplevelone(\indexone)} : \graphname{\hierarchygraphleft}{\maplevelone(\indextwo)} \partialmap \graphname{\hierarchygraphleft}{\maplevelone(\indexone)} \mid \chaindepthone \ge \indextwo > \indexone \ge 0)\)
Note, that \(\overline{\hierarchygraphleft}_{\downarrow \maplevelone}:=[\graphname{\hierarchygraphleft}{\maplevelone(\chaindepthone)}, 
\ldots,
\graphname{\hierarchygraphleft}{\maplevelone(0)}]\) is just a shorthand for the defining statement: \((\overline{\hierarchygraphleft}_{\downarrow \maplevelone})_\indexone := \graphname{\hierarchygraphleft}{\maplevelone(\indexone)}\) for all \(\chaindepthone \geq \indexone \geq 0\). 

The reduct morphism \((\overline{id}^{\chainname{\hierarchygraphleft}}_{\downarrow\maplevelone},\maplevelone) : \chainname[\downarrow\maplevelone]{\hierarchygraphleft} \to \chainname{\hierarchygraphleft}\) is a level-wise identity and just embeds an inclusion chain of length \(\chaindepthone+1\) into an inclusion chain of length \(\chaindepthtwo+1\), i.e.,
$
\overline{id}^{\chainname{\hierarchygraphleft}}_{\downarrow\maplevelone} = (id_{\maplevelone(\indexone)}: \graphname{\hierarchygraphleft}{\maplevelone(\indexone)} \to \graphname{\hierarchygraphleft}{\maplevelone(\indexone)} \mid \indexone \in [\chaindepthone] ).
$
In the pushout step (1) we will construct a pushout of inclusion chains of equal length and obtain a chain \(\chainname[\downarrow\maplevelone]{\hierarchygraphmiddle} := \chain[\downarrow\maplevelone]{\hierarchygraphmiddle}{\chaindepthone}\) with
\(\overline{\hierarchygraphmiddle}_{\downarrow\maplevelone} = [\graphname{\hierarchygraphmiddle}{\maplevelone(\chaindepthone)}, \graphname{\hierarchygraphmiddle}{\maplevelone(\chaindepthone-1)},\ldots,\graphname{\hierarchygraphmiddle}{\maplevelone(1)},\graphname{\hierarchygraphmiddle}{\maplevelone(0)=0}]\)
and
\(\typemorph[\hierarchygraphmiddle]{}{}_{\downarrow\maplevelone} = (\typemorph[\hierarchygraphmiddle]{\maplevelone(\indextwo)}{\maplevelone(\indexone)} : \graphname{\hierarchygraphmiddle}{\maplevelone(\indextwo)} \partialmap \graphname{\hierarchygraphmiddle}{\maplevelone(\indexone)} \mid \chaindepthone \ge \indextwo > \indexone \ge 0)\).

In the pushout step (2) we will fill the gaps in \(\chainname[\downarrow\maplevelone]{\hierarchygraphmiddle}\) with the corresponding untouched graphs from the original inclusion chain \(\chainname{\hierarchygraphleft}\).

\vspace{-2ex}
\subsubsection{Pushouts of graphs for inclusion graph homomorphisms.}
\label{subsubsec:pushout-inclusion-graph-homomorphisms}
Our constructions and proofs rely on the standard construction of pushouts in \cat{Graph}  for a span of an inclusion graph homomorphism \(\morphone : \graphname{\examplegraphone}{} \hookrightarrow \graphname{\examplegraphtwo}{}\) and an arbitrary graph homomorphism \(\morphtwo : \graphname{\examplegraphone}{} \to \graphname{\examplegraphthree}{}\) where we assume that \(\graphname{\examplegraphtwo}{}\) and \(\graphname{\examplegraphthree}{}\) are disjoint.
The pushout \(\graphname{\examplegraphfour}{}\) is given by \(\graphname[\graphnodes]{\examplegraphfour}{} := \graphname[\graphnodes]{\examplegraphthree}{} \cup \graphname[\graphnodes]{\examplegraphtwo}{} \setminus \graphname[\graphnodes]{\examplegraphone}{}\), \(\graphname[\grapharrows]{\examplegraphfour}{} := \graphname[\grapharrows]{\examplegraphthree}{} \cup \graphname[\grapharrows]{\examplegraphtwo}{} \setminus \graphname[\grapharrows]{\examplegraphone}{}\) and
\(\source[\examplegraphfour](e) := 
\source[\examplegraphthree](e)\), if  \(e \in \graphname[\grapharrows]{\examplegraphthree}{}\), and 
\(\source[\examplegraphfour](e) :=\morphtwo^\grapharrows(\source[\examplegraphtwo](e))\), if \(e \in \graphname[\grapharrows]{\examplegraphtwo}{} \setminus \graphname[\grapharrows]{\examplegraphone}{}\).
\(\target[\examplegraphfour]\) is defined analogously.
\(\morphone^*: \graphname{\examplegraphthree}{} \hookrightarrow \graphname{\examplegraphfour}{}\) is an inclusion graph homomorphism by construction and \(\morphtwo^*: \graphname{\examplegraphtwo}{} \to \graphname{\examplegraphfour}{}\) is defined for \(X\in\{\grapharrows,\graphnodes\}\) by \(\morphtwo^{*X}(v) := \morphtwo^X(v)\), if \(v \in \graphname[X]{\examplegraphone}{}\) and 
\(\morphtwo^{*X}(v) := v\) , if \(v \in \graphname[X]{\examplegraphtwo}{} \setminus \graphname[X]{\examplegraphone}{}\).

The pair \(\graphname{\examplegraphone}{} \setminus \graphname{\examplegraphtwo}{} := (\graphname[\graphnodes]{\examplegraphtwo}{} \setminus \graphname[\graphnodes]{\examplegraphone}{}, \graphname[\grapharrows]{\examplegraphtwo}{} \setminus \graphname[\grapharrows]{\examplegraphone}{})\) of subsets of nodes and arrows of \(\graphname{\examplegraphtwo}{}\) is, in general, not establishing a subgraph of \(\graphname{\examplegraphtwo}{}\).
We will nevertheless use the notation \(\graphname{\examplegraphfour}{} = \graphname{\examplegraphthree}{} + \graphname{\examplegraphtwo}{} \setminus \graphname{\examplegraphone}{}\) to indicate that \(\graphname{\examplegraphfour}{}\) is constructed as described above.
\(\morphtwo^*\) can be described then as a sum of two parallel pairs of mappings
\begin{equation}\label{eq:sum-mappings}
\morphtwo^*= \morphtwo+ id_{\graphname{\examplegraphtwo}{}\setminus\graphname{\examplegraphone}{}} := (\morphtwo^\graphnodes + id_{\graphname[\graphnodes]{\examplegraphtwo}{}\setminus\graphname[\graphnodes]{\examplegraphone}{}}, \morphtwo^\grapharrows + id_{\graphname[\grapharrows]{\examplegraphtwo}{}\setminus\graphname[\grapharrows]{\examplegraphone}{}})
\end{equation}

\subsubsection{Pushout for inclusion chains with equal depth.}
\label{subsubsec:pushout-chain-equal-depth}

We consider now the span
\vspace{-2.5ex}
\begin{center}
	\begin{tikzpicture}[on grid,node distance=35mm]
	
	\node[el-math] (sf)					{\chainname[\downarrow\maplevelone]{\hierarchygraphleft}};
	\node[el-math] (l)	[right of=sf]	{\chainname{\rulegraphleft}};
	\node[el-math] (i)	[right of=l]	{\chainname{\rulegraphmiddle}};
	
	\draw[map]		(l) to node[la-math,above]			(mu1)	{(\mapruletohierarchyleft,id_{[\chaindepthone]})}											(sf);
	\draw[incmapl]	(l) to node[la-math,above]	(in)	{(\maprulelefttomiddle,id_{[\chaindepthone]})}	(i);
	
	\end{tikzpicture}
	\vspace{-1ex}
\end{center}
of reduct morphisms in \(\cat{Chain}\) (see Fig.~\ref{fig:two-po-construct-d}). 
For each level \(\indexone \in [\chaindepthone]\)  we construct the corresponding pushout of graph homomorphisms.
%
%
%
%
This ensures, especially, 
\begin{equation}\label{eq:commutativity-po-step1-levelwise}
\maprulelefttomiddle_{\indexone};\mapruletohierarchymiddle_\indexone = \mapruletohierarchyleft_{\indexone};\maphierarchylefttomiddle_{\maplevelone(\indexone)} \quad\mbox{for all}\quad
\indexone \in [\chaindepthone].
\end{equation}
\vspace{-3ex}
\begin{figure}[ht!]
\begin{center}
	\begin{tikzpicture}[on grid,node distance=30mm]

	\node[el-math] (l0)		            {\graphname{\rulegraphleft}{0}=\graphname{\rulegraphleft}{}};
	\node[el-math] (i0)		[right=35mm  of l0]						{\graphname{\rulegraphmiddle}{0}=\graphname{\rulegraphmiddle}{}};
	\node[el-math] (li)		[below of=l0]						{\graphname{\rulegraphleft}{\indexone}};
	\node[el-math] (ii)		[right=35mm  of li]						{\graphname{\rulegraphmiddle}{\indexone}};
	\node[el-math] (s0)		[below right=15mm and 16mm of l0]	{\graphname{\hierarchygraphleft}{0}=\graphname{\hierarchygraphleft}{}};
	\node[el-math] (d0)		[right=35mm  of s0]						{\graphname{\hierarchygraphmiddle}{0}=\graphname{\hierarchygraphmiddle}{}= \graphname{\hierarchygraphleft}{} + \graphname{\rulegraphmiddle}{} \setminus \graphname{\rulegraphleft}{}};
	\node[el-math] (sfi)	[below of=s0]						{\graphname{\hierarchygraphleft}{\maplevelone(\indexone)}};
	\node[el-math] (dfi)	[right=35mm  of sfi]						{\graphname{\hierarchygraphmiddle}{\maplevelone(\indexone)}= \graphname{\hierarchygraphleft}{\maplevelone(\indexone)} + \graphname{\rulegraphmiddle}{\indexone} \setminus \graphname{\rulegraphleft}{\indexone}};
	\node[el-math] (po1)	[below right=8mm and 25mm of l0]	{\po};
	\node[el-math] (po2)	[below of=po1]						{\po};
	
	\draw[incmapl]	(l0) to node[la-math]				(mg0i0)	{\maprulelefttomiddle_0=\maprulelefttomiddle}	(i0);
	\draw[incmapl]	(li) to node[la-math,pos=.7]		(mgiii)	{\maprulelefttomiddle_\indexone}				(ii);
	\draw[incmapr]	(li) to node[la-math]				(mgig0)	{\typemorph[\rulegraphleft]{\indexone}{0}}		(l0);
	\draw[incmapr]	(ii) to node[la-math,pos=.25,right]	(miii0)	{\typemorph[\rulegraphmiddle]{\indexone}{0}}	(i0);
	
	\draw[incmapl]	(s0) to node[la-math,pos=.4,below]	(ms0d0)	{\maphierarchylefttomiddle_{0}=\maphierarchylefttomiddle}		(d0);
	\draw[incmapl]	(sfi) to node[la-math,below]		(mkfpf)	{\maphierarchylefttomiddle_{\maplevelone(\indexone)}}			(dfi);
	\draw[incmapr]	(sfi) to node[la-math,pos=.7]		(mkfs0)	{\typemorph[\hierarchygraphleft]{\maplevelone(\indexone)}{0}}	(s0);
	\draw[mapdots]	(dfi) to node[la-math,right]		(mpfd0)	{\typemorph[\hierarchygraphmiddle]{\maplevelone(\indexone)}{0} (=\typemorph[\hierarchygraphleft]{\maplevelone(\indexone)}{0}+\typemorph[\rulegraphmiddle]{\indexone}{0\downarrow\graphname{\rulegraphmiddle}{\indexone}\setminus\graphname{\rulegraphleft}{\indexone}})}	(d0);
	
	\draw[map]	(l0) to node[la-math]	(mg0s0)	{\mapruletohierarchyleft_0}			(s0);
	\draw[map]	(i0) to node[la-math]	(mi0d0)	{\mapruletohierarchymiddle_0}			(d0);
	\draw[map]	(li) to node[la-math]	(mgikf)	{\mapruletohierarchyleft_\indexone}	(sfi);
	\draw[map]	(ii) to node[la-math]	(miipf)	{\mapruletohierarchymiddle_\indexone}	(dfi);
	
	\end{tikzpicture}
\end{center}
	\vspace{-4ex}
	\caption{Level-wise pushout construction}
	\label{fig:level-wise-pushout}
	\vspace{-3ex}
\end{figure}

We look at an arbitrary level \(\chaindepthone \geq \indexone \geq 1\) together with the base level \(0\) (see Fig.~\ref{fig:level-wise-pushout}). We get a cube where the top and bottom square are pushouts by construction. In addition, the left and back square are pullbacks since \((\mapruletohierarchyleft,id_{[\chaindepthone]})\) and \((\maprulelefttomiddle,id_{[\chaindepthone]})\), respectively, are reduct morphisms.
We get a unique graph homomorphism \(\typemorph[\hierarchygraphmiddle]{\maplevelone(\indexone)}{0}: \graphname{\hierarchygraphmiddle}{\maplevelone(\indexone)} \to \graphname{\hierarchygraphmiddle}{}\) that makes the cube commute. By the uniqueness of mediating morphisms and the fact that the top pushout square has the Van Kampen property (see \cite{ehrig2006fundamentals,WolterK15}), we can conclude that the front and the the right square are pullbacks as well.
That the back square is a pullback means nothing but \(\graphname{\rulegraphleft}{\indexone} = \graphname{\rulegraphleft}{}\cap\graphname{\rulegraphmiddle}{\indexone}\). This entails \(\graphname{\rulegraphmiddle}{\indexone}\setminus\graphname{\rulegraphleft}{\indexone} 
\subseteq \graphname{\rulegraphmiddle}{}\setminus\graphname{\rulegraphleft}{}\)  thus \(\typemorph[\hierarchygraphmiddle]{\maplevelone(\indexone)}{0}\) turns out to be the sum of the two inclusions \(\typemorph[\hierarchygraphleft]{\maplevelone(\indexone)}{0}: \graphname{\hierarchygraphleft}{\maplevelone(\indexone)} \hookrightarrow\graphname{\hierarchygraphleft}{}\) and \(\typemorph[\rulegraphmiddle]{\indexone}{0\downarrow\graphname{\rulegraphmiddle}{\indexone}\setminus\graphname{\rulegraphleft}{\indexone}}: \graphname{\rulegraphmiddle}{\indexone}\setminus\graphname{\rulegraphleft}{\indexone}\hookrightarrow\graphname{\rulegraphmiddle}{}\setminus\graphname{\rulegraphleft}{}\) and is therefore an inclusion itself.

The sequence \([\graphname{\hierarchygraphmiddle}{\maplevelone(\chaindepthone)}, \graphname{\hierarchygraphmiddle}{\maplevelone(\chaindepthone-1)}, \dots, \graphname{\hierarchygraphmiddle}{\maplevelone(1)}, \graphname{\hierarchygraphmiddle}{0}]\) of subgraphs of \(\graphname{\hierarchygraphmiddle}{}=\graphname{\hierarchygraphmiddle}{0}\) defines the intended inclusion chain \(\chainname[\downarrow\maplevelone]{\hierarchygraphmiddle}\).
Since the front and right squares in Fig.~\ref{fig:level-wise-pushout} are pullbacks, Lemma \ref{lem:inclusion-chains-establish-chain-morphism} ensures that the family  \(\maphierarchylefttomiddle_{\downarrow\maplevelone}=(\maphierarchylefttomiddle_{\maplevelone(\indexone)}: \graphname{\hierarchygraphleft}{\maplevelone(\indexone)} \hookrightarrow \graphname{\hierarchygraphmiddle}{\maplevelone(\indexone)}  \mid \indexone \in [\chaindepthone])\) of inclusion graph homomorphisms constitutes a reduct morphism \((\maphierarchylefttomiddle_{\downarrow\maplevelone},id_{[\chaindepthone]}): \chainname[\downarrow\maplevelone]{\hierarchygraphleft} \rightarrow \chainname[\downarrow\maplevelone]{\hierarchygraphmiddle}\) while the family   \(\mapruletohierarchymiddle=(\mapruletohierarchymiddle_{\indexone}: \graphname{\rulegraphmiddle}{\indexone} \rightarrow \graphname{\hierarchygraphmiddle}{\maplevelone(\indexone)}  \mid \indexone \in [\chaindepthone])\) of graph homomorphisms constitutes a reduct morphism \((\mapruletohierarchymiddle,id_{[\chaindepthone]}): \chainname{\rulegraphmiddle} \rightarrow \chainname[\downarrow\maplevelone]{\hierarchygraphmiddle}\). Finally, equation \myref{eq:commutativity-po-step1-levelwise} ensures that the resulting square (1) of reduct morphisms in Fig.~\ref{fig:two-po-construct-d} commutes.
The proof that we have constructed a pushout in \cat{Chain} is given in \cite{wolter2019chains}.

\vspace{-.5ex}
\begin{remark}[Only one pushout]
	\label{rem:only-one-pushout}
	\(\maphierarchylefttomiddle_{\maplevelone(\indexone)}\) and \(\mapruletohierarchymiddle_{\indexone}\) are jointly surjective for all \(\chaindepthone \geq \indexone \geq 1\) thus  we can describe \(\graphname{\hierarchygraphmiddle}{\maplevelone(\indexone)}\) as the union \(\graphname{\hierarchygraphmiddle}{\maplevelone(\indexone)}= \maphierarchylefttomiddle(\graphname{\hierarchygraphleft}{\maplevelone(\indexone)}) \cup \mapruletohierarchymiddle(\graphname{\rulegraphmiddle}{\indexone})\). 
	Hence in practice, there is no need for an explicit construction of pushouts at all the levels \(\chaindepthone \geq \indexone \geq 1\); these are all constructed implicitly by the pushout construction at level $0$. 
\end{remark}

\vspace{-3ex}
\subsubsection{Pushout by chain extension.}
\label{subsubsec:pushout-by-extension}

To obtain an inclusion chain \(\chainname{\hierarchygraphmiddle}\) of length \(\chaindepthtwo+1\), we fill the gaps in \(\chainname[\downarrow\maplevelone]{\hierarchygraphmiddle}\)
by corresponding subgraphs of \(\hierarchygraphleft\):
\(\graphname{\hierarchygraphmiddle}{\exampleindexone} :=\graphname{\hierarchygraphmiddle}{\exampleindexone}\) if \(\exampleindexone \in \maplevelone([\chaindepthone])\) and \(\graphname{\hierarchygraphmiddle}{\exampleindexone} := \graphname{\hierarchygraphleft}{\exampleindexone}\) if \(\exampleindexone \in [\chaindepthtwo] \setminus \maplevelone([\chaindepthone])\)
and obtain the intended inclusion chain \(\chainname{\hierarchygraphmiddle} = \chain{\hierarchygraphmiddle}{\chaindepthtwo}\).
The family \(\overline{id}^{\chainname{\hierarchygraphmiddle}}_{\downarrow\maplevelone} = (id_{\graphname{\hierarchygraphmiddle}{\maplevelone(\indexone)}}: \graphname{\hierarchygraphmiddle}{\maplevelone(\indexone)} \to \graphname{\hierarchygraphmiddle}{\maplevelone(\indexone)} \mid \indexone \in [\chaindepthone])\) of identities defines trivially a reduct morphism \((\overline{id}^{\chainname{\hierarchygraphmiddle}}_{\downarrow\maplevelone},\maplevelone) : \chainname[\downarrow\maplevelone]{\hierarchygraphmiddle} \to \chainname{\hierarchygraphmiddle}\).
One can  show that the family \(\maphierarchylefttomiddle = (\maphierarchylefttomiddle_{\exampleindexone}: \graphname{\hierarchygraphleft}{\exampleindexone} \to \graphname{\hierarchygraphmiddle}{\exampleindexone} \mid a \in [\chaindepthtwo])\) of graph homomorphisms defined by
\begin{equation*}
\maphierarchylefttomiddle_{\exampleindexone} := \begin{cases}
\maphierarchylefttomiddle_{\exampleindexone} : \graphname{\hierarchygraphleft}{\exampleindexone} \hookrightarrow \graphname{\hierarchygraphmiddle}{\exampleindexone} &\mbox{if } a \in \maplevelone([\chaindepthone])\\
id_{\graphname{\hierarchygraphleft}{\exampleindexone}} : \graphname{\hierarchygraphleft}{\exampleindexone} \to \graphname{\hierarchygraphmiddle}{\exampleindexone} = \graphname{\hierarchygraphleft}{\exampleindexone} &\mbox{if } a \in [\chaindepthtwo] \setminus \maplevelone([\chaindepthone])
\end{cases}
\end{equation*}
establishes a reduct morphism \((\maphierarchylefttomiddle, id_{[\chaindepthtwo]}) : \chainname{\hierarchygraphleft} \to \chainname{\hierarchygraphmiddle}\).
The two reduct morphisms \((\overline{id}^{\chainname{\hierarchygraphmiddle}}_{\downarrow\maplevelone},\maplevelone) : \chainname[\downarrow\maplevelone]{\hierarchygraphmiddle} \to \chainname{\hierarchygraphmiddle}\) and \((\maphierarchylefttomiddle, id_{[\chaindepthtwo]}) : \chainname{\hierarchygraphleft} \to \chainname{\hierarchygraphmiddle}\) establish square (2) in  Fig.~\ref{fig:two-po-construct-d} that commutes trivially. In\cite{wolter2019chains} it is shown that square (2) is also a pushout in \cat{Chain}.




\subsubsection{Pullback complement.}
\label{subsubsec:pullbacks-category-chain}

%
We construct the reduct of \(\chainname{\hierarchygraphmiddle} = \chain{\hierarchygraphmiddle}{\chaindepthtwo}\) along \(\maphierarchyrighttomiddle:\graphname{\hierarchygraphright}{}\hookrightarrow\graphname{\hierarchygraphmiddle}{}\) and \(id_{[\chaindepthtwo]}\) by level-wise intersection (pullback) for all \(\chaindepthone \geq \indexone \geq 1\) (see the pullback square below). Due to Lemma \ref{lem:inclusion-chains-establish-chain-morphism}, we obtain, in such a way, an inclusion chain \(\chainname{\hierarchygraphright} = \chain{\hierarchygraphright}{\chaindepthtwo}\)  together with a reduct morphism \((\maphierarchyrighttomiddle,id_{[\chaindepthtwo]}):\chainname{\hierarchygraphright}\to\chainname{\hierarchygraphmiddle}\).
The multilevel typing of \(\chainname{\hierarchygraphright}{}\) is simply borrowed from \(\chainname{\hierarchygraphmiddle}\), that is, we define (see Fig.~\ref{fig:pullback-complement-step})
\begin{equation}\label{eq:borrowing-of-typing}
(\chainmorph{\hierarchygraphright},id_{[\chaindepthtwo]}) := (\maphierarchyrighttomiddle,id_{[\chaindepthtwo]}); (\chainmorph{\hierarchygraphmiddle},id_{[\chaindepthtwo]})
\end{equation}
and this gives us trivially the intended type compatibility of \((\maphierarchyrighttomiddle,id_{[\chaindepthtwo]})\).
The typing chain morphism \((\mapruletohierarchyright,\maplevelone) : \chainname{\rulegraphright} \to \chainname{\hierarchygraphright}\) 
with \(\mapruletohierarchyright = (\mapruletohierarchyright_\indexone : \graphname{\rulegraphright}{\indexone} \to \graphname{\hierarchygraphright}{\maplevelone(\indexone)} \mid \indexone \in [\chaindepthone])\) such that
\vspace{-1ex}
\begin{equation}\label{eq:chainmorph-r-to-t}
(\maprulerighttomiddle,id_{[\chaindepthone]});(\mapruletohierarchymiddle,\maplevelone) = (\mapruletohierarchyright,\maplevelone);(\maphierarchyrighttomiddle,id_{[\chaindepthtwo]})
\end{equation}
is simply given by pullback composition and decomposition in \cat{Graph}: For each \(\chaindepthone \geq \indexone \geq 1\) we consider the following incomplete cube on the right-hand side:
\vspace{-1ex}
\begin{center}
	\begin{tikzpicture}[on grid,node distance=26mm]

	\node[el-math] (d0a)	{\graphname{\hierarchygraphmiddle}{0} = \graphname{\hierarchygraphmiddle}{}};
    \node[el-math] (t0a)    [right=30mm of d0a]	{\graphname{\hierarchygraphright}{0} = \graphname{\hierarchygraphright}{}};
    \node[el-math] (dia)	[below of=d0a]		{\graphname{\hierarchygraphmiddle}{\indexone}};
    \node[el-math] (tia)	[below of=t0a]		{\graphname{\hierarchygraphright}{\indexone}};

    \draw[incmapr]	(t0a) to node[la-math,above]		(mt0d0a)	{\maphierarchyrighttomiddle_{0} = \maphierarchyrighttomiddle}	(d0a);
    \draw[incmapr]	(tia) to node[la-math,above]		(mtidia)	{\maphierarchyrighttomiddle_{\indexone}}			(dia);
    \draw[incmapl]	(dia) to node[la-math]				(mdid0a)	{\typemorph[\hierarchygraphmiddle]{\indexone}{0}}	(d0a);
    \draw[incmapl]	(tia) to node[la-math,right]				(mtit0a)	{\typemorph[\hierarchygraphright]{\indexone}{0}}	(t0a);

    \node[el-math] (pba)	[below=14mm of mt0d0a]	{\pb};
	
	\node[el-math] (i0)	    [right=30mm of t0a]		{\graphname{\rulegraphmiddle}{0} = \graphname{\rulegraphmiddle}{}};
	\node[el-math] (r0)		[right of=i0]						{\graphname{\rulegraphright}{0} = \graphname{\rulegraphright}{}};
	\node[el-math] (ii)		[below of=i0]						{\graphname{\rulegraphmiddle}{\indexone}};
	\node[el-math] (ri)		[right of=ii]						{\graphname{\rulegraphright}{\indexone}};
	\node[el-math] (d0)		[below right=15mm and 15mm of i0]	{\graphname{\hierarchygraphmiddle}{0} = \graphname{\hierarchygraphmiddle}{}};
	\node[el-math] (t0)		[right of=d0]						{\graphname{\hierarchygraphright}{0} = \graphname{\hierarchygraphright}{}};
	\node[el-math] (dfi)	[below of=d0]						{\graphname{\hierarchygraphmiddle}{\maplevelone(\indexone)}};
	\node[el-math] (tfi)	[right of=dfi]						{\graphname{\hierarchygraphright}{\maplevelone(\indexone)}};
	
	\draw[incmapr] (r0) to	node[la-math,above]			(mr0i0)	{\maprulerighttomiddle_0 = \maprulerighttomiddle}	(i0);
	\draw[incmapr] (ri) to	node[la-math,above,pos=.7]	(mriii)	{\maprulerighttomiddle_\indexone}					(ii);
	\draw[incmapr] (ii) to	node[la-math,pos=.3]		(msj0)	{\typemorph[\rulegraphmiddle]{\indexone}{0}}		(i0);
	\draw[incmapr] (ri) to	node[la-math,right,pos=.21]		(mdj0)	{\typemorph[\rulegraphright]{\indexone}{0}}			(r0);
	
	\draw[incmapr] (t0)  to	node[la-math,above,pos=.55]	(mt0d0)		{\maphierarchyrighttomiddle_0 = \maphierarchyrighttomiddle}		(d0);
	\draw[incmapr] (tfi) to	node[la-math,above]			(mtfidfi)	{\maphierarchyrighttomiddle_{\maplevelone(\indexone)}}			(dfi);
	\draw[incmapr] (dfi) to	node[la-math,right,pos=.34]		(mdfid0)	{\typemorph[\hierarchygraphmiddle]{\maplevelone(\indexone)}{0}}	(d0);
	\draw[incmapr] (tfi) to	node[la-math,right,pos=.7]	(mtfit0)	{\typemorph[\hierarchygraphright]{\maplevelone(\indexone)}{0}}	(t0);
	
	\draw[map]	 	(i0) to	node[la-math]	(mi0d0)		{\mapruletohierarchymiddle_0 = \mapruletohierarchymiddle}	(d0);
	\draw[map] 	 	(r0) to	node[la-math]	(mr0t0)		{\mapruletohierarchyright_0 = \mapruletohierarchyright}		(t0);
	\draw[map] 		(ii) to	node[la-math,below,pos=.4]	(miidfi)	{\mapruletohierarchymiddle_\indexone}						(dfi);
	\draw[mapdots] 	(ri) to	node[la-math]	(mritfi)	{\mapruletohierarchyright_\indexone}						(tfi);
	
	\end{tikzpicture}
	\vspace{-1ex}
\end{center}

The back square, the left square as well as the front square are pullbacks since \((\maprulerighttomiddle,id_{[\chaindepthone]})\),
\((\mapruletohierarchymiddle,\maplevelone)\) and \((\maphierarchyrighttomiddle,id_{[\chaindepthtwo]})\), respectively, are reduct morphisms.
The top square is constructed as a pullback complement.
The diagonal square from \(\typemorph[\rulegraphright]{\indexone}{0}\) to \(\typemorph[\hierarchygraphmiddle]{\maplevelone(\indexone)}{0}\) is a pullback due to the composition of the back pullback and the left pullback.
The decomposition of this diagonal pullback w.r.t. the front pullback gives us \(\mapruletohierarchyright_\indexone : \graphname{\rulegraphright}{\indexone} \to \graphname{\hierarchygraphright}{\indexone}\) making the cube, and especially the bottom square, commute and making the right square to a pullback as well.

According to Lemma~\ref{lem:inclusion-chains-establish-chain-morphism} the family \(\mapruletohierarchyright = (\mapruletohierarchyright_\indexone : \graphname{\rulegraphright}{\indexone} \to \graphname{\hierarchygraphright}{\maplevelone(\indexone)} \mid \indexone \in [\chaindepthone])\) of graph homomorphisms defines a reduct morphism \((\mapruletohierarchyright,\maplevelone) : \chainname{\rulegraphright} \to \chainname{\hierarchygraphright}\) where condition~\myref{eq:chainmorph-r-to-t} is simply satisfied by construction. 
Finally,  \((\mapruletohierarchyright,\maplevelone)\) is also type compatible since conditions \myref{eq:borrowing-of-typing} and \myref{eq:chainmorph-r-to-t} ensure that the roof square in Fig.~\ref{fig:pullback-complement-step} commutes.

\begin{figure}
  \centering
  \begin{minipage}[t]{.35\linewidth}
    \centering
    \includestandalone[scale=.6]{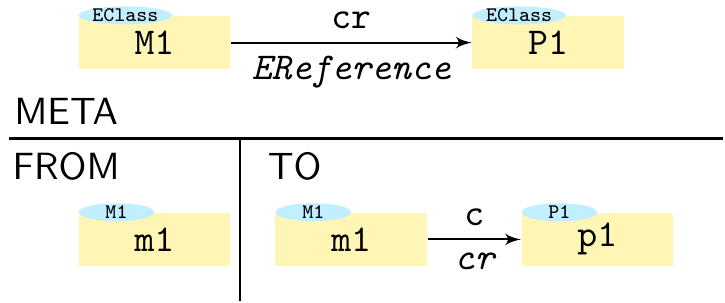}
  \end{minipage}
  \quad
  \begin{minipage}[t]{.2\linewidth}
    \centering
    \includestandalone[scale=.7]{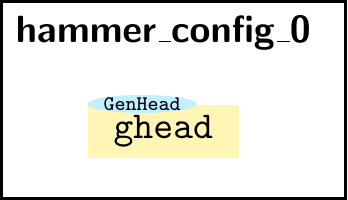}
  \end{minipage}
  \quad
  \begin{minipage}[t]{.35\linewidth}
    \centering
    \includestandalone[scale=.7]{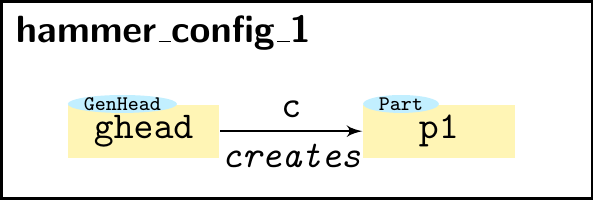}
  \end{minipage}
  \caption{Plain version of \textit{CreatePart} and its application}
  \label{fig:create-part-modified-and-application}
\end{figure}

\begin{example}
\label{ex:rule-application}
To present a non-trivial rule application for our  example, we discuss the undesired application of the plain version of rule \textit{CreatePart} (see Fig.~\ref{fig:create-part-modified-inclusion-chains}), mentioned in Example \ref{ex:match+rule-application}, for a  state of the hammer configuration with only one node \elementname{ghead}, as shown in \elementname{hammer\_config\_0} in Fig.~\ref{fig:create-part-modified-and-application}. So, we have \(\maplevelone:[1]\to[2]\), with \(\maplevelone(0)=0, \maplevelone(1)=1\), and the ``undesired match'' of \graphname{\rulegraph}{1} =
(\begin{tikzpicture}[baseline=-1mm,node distance=12mm]
\node[el-math](m1){\elementname{M1}};
\node[el-math](p1)[right of=m1]{\elementname{P1}};
\draw[map](m1)to	node[la-math,above](rm){\elementname{cr}}(p1);
\end{tikzpicture}) 
with \graphname{\hierarchygraph}{1} = \elementname{generic\_plant}  =
(\begin{tikzpicture}[baseline=-1mm,node distance=26mm]
\node[el-math](m1){\elementname{Machine}};
\node[el-math](p1)[right of=m1]{\elementname{Part}};
\draw[map](m1)to	node[la-math,above](rm){\elementname{creates}}(p1);
\end{tikzpicture})
together with the trivial match of the left-hand side \(\rulegraphleft=(\,\elementname{m1}\,)\) of the rule with \(\elementname{hammer\_config\_0} = (\,\elementname{ghead}\,)\). The resulting inclusion chains \chainname{\hierarchygraphleft}, \chainname{\rulegraphleft}, \chainname{\rulegraphright} and two reduct morphisms between them are depicted in Fig.~\ref{fig:create-part-modified-inclusion-chains}. Note, that the ellipse and cursive labels indicate here the corresponding typing chain morphisms \((\chainmorph{\hierarchygraphleft},id_{[2]})\), \((\chainmorph{\rulegraphleft},id_{[1]})\) and \((\chainmorph{\rulegraphright},id_{[1]})\), respectively.
    \begin{figure}
    	\begin{center}
    		\includestandalone[width=\linewidth]{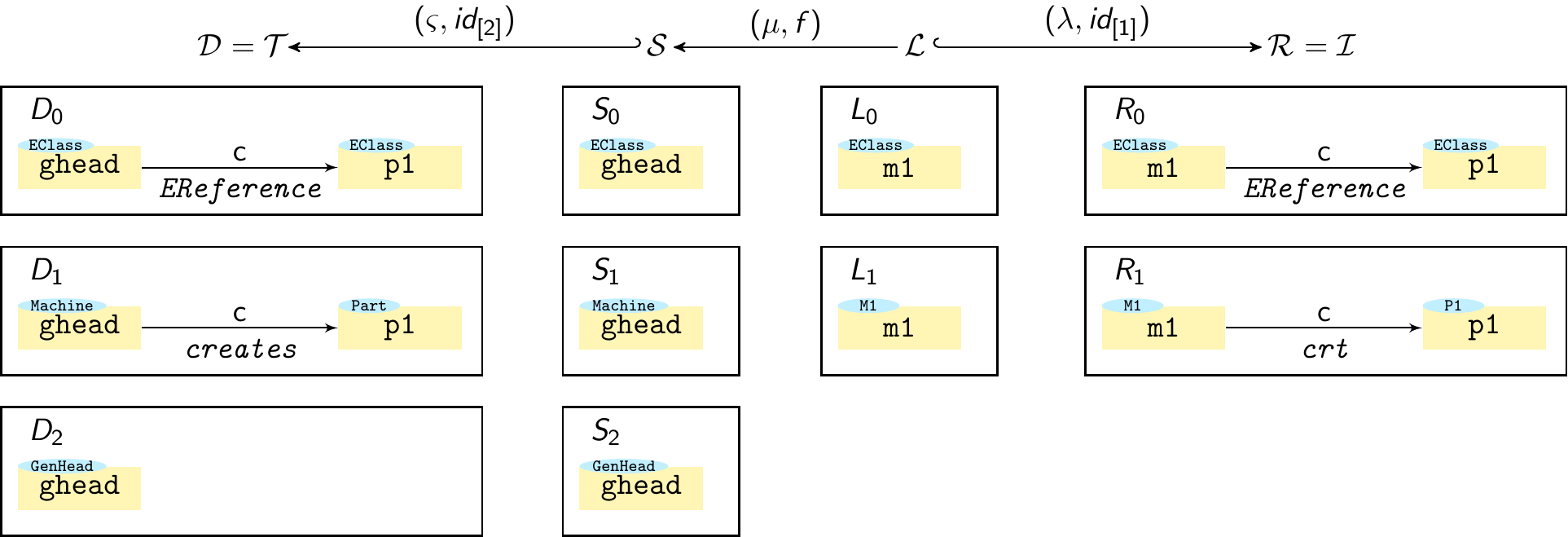}
    		\caption{Inclusion chains for the plain version of \textit{CreatePart}}
    		\label{fig:create-part-modified-inclusion-chains}
    	\end{center}
    	\vspace{-6ex}
    \end{figure}
    
For the two levels in \(\maplevelone([1])=\{0,1\}\subset [2]\) we construct the pushouts \graphname{\hierarchygraphmiddle}{0} and \graphname{\hierarchygraphmiddle}{1} while \graphname{\hierarchygraphmiddle}{2} is just taken as \graphname{\hierarchygraphleft}{2}. The lowest level in \chainname{D}, where the new elements \elementname{p1} and \elementname{c} appear, is level \(1\) thus the constructed direct types of \elementname{p1} and \elementname{c} are \elementname{Part} and \elementname{creates}, resp., as shown in \elementname{hammer\_config\_1} in Fig.~\ref{fig:create-part-modified-and-application}.
\end{example}

\section{Conclusions, related and future work}
\label{sec:conclusion}

\textbf{Conclusion.} Multilevel modeling offers more flexibility on top of traditional modeling techniques by supporting an unlimited number of abstraction levels. 
Our approach to multilevel modeling enhances reusability of concepts and their behaviour by allowing the definition of flexible transformation rules which are applicable to different hierarchies with a variable number of levels.
In this paper, we have presented a formalization of these flexible and reusable transformation rules based on graph transformations.
We represent multilevel models by multilevel typed graphs whose manipulation and transformation are carried out by multilevel typed graph transformation rules.
These rules are cospans of three graphs and two inclusion graph homomorphisms where the three graphs are multilevel typed over a common typing chain.
As these rules are represented as cospans, their application is carried out by a pushout and a final pullback complement construction for the underlying graphs in the category \cat{Graph}.
We have identified type compatibility conditions, for rules and their matches, which are crucial for rule applications. Moreover, we have shown that typed graph transformations can be generalized to multilevel typed graph transformations improving preciseness, flexibility and reusability of transformation rules. 

\noindent
\textbf{Related work.}
The theory and practise of graph transformations are well-studied, and the concept of model transformations applied to MLM is not novel.
Earlier works in the area have worked in the extension of pre-existing model transformation languages to be able to manipulate multilevel models and model hierarchies.
In~\cite{atkinson2015enhancing}, the authors adapt ATL~\cite{JouaultABK08} to manipulate multilevel models built with the Melanee tool~\cite{atkinson2016melanee}.
In a similar manner,~\cite{delara2015dsmml} proposes the adaptation of ETL~\cite{KolovosPP08} and other languages from the Epsilon family~\cite{Epsilon} for the application of model transformation rules into multilevel hierarchies created with MetaDepth~\cite{delara2010deep}.
These works, however, tackle the problem from the practical point of view.
That is, how to reuse mature off-the-shelf tools for model transformation in the context of MLM, via the manipulation of a ``flattened'' representation of the hierarchy to emulate multilevel transformations.
Our approach, on the contrary, has been developed from scratch with a multilevel setting in mind, and we believe it can be further extended to tackle all scenarios considered by other approaches.
Therefore, to the best of our knowledge, there are no formal treatments of multilevel typed graph transformations in the literature except for our previous works~\cite{wolter2019chains,macias19jlamp,macias2016emisa} (see Sect.~4 in \cite{wolter2019chains}).
Hence, we consider our approach the first approximation to formally address the challenges which come with multilevel modeling and multilevel model transformations.

Common for our work and \cite{delara2020mlmpl}  is that the concepts of typing chains, multilevel typed graphs and multilevel models  originate in \cite{rossini2014formalisation}. 
However, \cite{delara2020mlmpl} presents partial morphisms as spans of total morphisms and does not use the composition of those spans explicitly.
Wrt. typing chains, a multilevel model in \cite{delara2020mlmpl} is a sequence of graphs \([\graphname{\examplegraphone}{\chaindepthone}, \graphname{\examplegraphone}{\chaindepthone-1}, \dots, \graphname{\examplegraphone}{1}, \graphname{\examplegraphone}{0}]\) together with the subfamily \((\typemorph[\examplegraphone]{\indexone+1}{\indexone} : \graphname{\examplegraphone}{\indexone+1} \partialmap \graphname{\examplegraphone}{\indexone} \mid \chaindepthone \geq \indexone \geq 0 )\) of typing morphisms.

\noindent
\textbf{Future work.} 
Although it is trivial to see that the bottom square in the cube for the pullback complement step becomes a pullback for all \(\chaindepthone \geq \indexone \geq 1\), we leave it for future work to prove that we indeed have constructed a final pullback complement in \cat{Chain}.
A utilization of our theory to deal with coupled transformations \cite{mantz2015coevolving} in the setting of multilevel typed modelling is also desirable.
Furthermore, it would be interesting to investigate the category \cat{Chain} for its own; e.g., study its monomorphisms and epimorphisms, possible factorization systems, and the conditions for existence of general pushouts and pullbacks.

\bibliographystyle{plain}
\bibliography{bibliography}
\end{document}